\documentclass[twocolumn,twocolappendix]{openjournal}
\usepackage{xcolor}
\usepackage{graphicx}
\usepackage{amsfonts}
\usepackage{amssymb}
\usepackage{url}
\usepackage[breaklinks,colorlinks,citecolor=blue,linkcolor=blue,urlcolor=blue]{hyperref}
\usepackage{amsmath}

\def\Eu{ \mathfrak{H} } 
\def\la{\langle}
\def\ra{\rangle}

\begin{document}

\title{Towards Cosmography of the Local Universe}

\author{Julian Adamek}
\affiliation{Institut f\"ur Astrophysik, Universit\"at Z\"urich,\\Winterthurerstrasse 190, 8057 Z\"urich, Switzerland}

\author{Chris Clarkson}
\affiliation{School of Physical and Chemical Sciences, Queen Mary University of London,\\327 Mile End Road, London E1 4NS, UK\\}
\affiliation{Department of Physics \& Astronomy, University of the Western Cape,\\Cape Town 7535, South Africa}

\author{Ruth Durrer}
\affiliation{D\'epartement de Physique Th\'eorique, Universit\'e de Gen\`eve,\\24 quai Ernest-Ansermet, 1211 Gen\`eve 4, Switzerland}

\author{Asta Heinesen}
\affiliation{Universit\'e de Lyon 1 \& ENS de Lyon \& CNRS \& Centre de Recherche Astrophysique de Lyon UMR5574,\\69007 Lyon, France\\} 
\affiliation{Niels Bohr Institute,\\Blegdamsvej 17, 2100 Copenhagen, Denmark}

\author{Martin Kunz}
\affiliation{D\'epartement de Physique Th\'eorique, Universit\'e de Gen\`eve,\\24 quai Ernest-Ansermet, 1211 Gen\`eve 4, Switzerland}

\author{Hayley J.\ Macpherson}
\affiliation{Kavli Institute for Cosmological Physics, The University of Chicago, \\5640 South Ellis Avenue, Chicago, Illinois 60637, USA\\}
\affiliation{NASA Einstein Fellow\\}

\begin{abstract}
Anisotropies in the distance--redshift relation of cosmological sources are expected due to large-scale inhomogeneities in the local Universe. When the observed sources are tracing a large-scale matter flow
in a general spacetime geometry, the distance--redshift relation
with its anisotropies can be described with a geometrical prediction that generalises the well-known Friedmann-Lema\^itre-Robertson-Walker
result. Furthermore, it turns out that a finite set of multipole coefficients contain the full information about a finite-order truncation of the distance--redshift relation of a given observer. 
The multipoles of the distance--redshift relation are interesting new cosmological observables that have a direct physical interpretation in terms of kinematical quantities of the underlying matter flow. 
Using light cones extracted from $N$-body simulations we quantify the anisotropies expected in a $\Lambda$ cold dark matter cosmology by running a Markov chain Monte Carlo analysis on the observed data. In this observational approach the survey selection implements an implicit smoothing scale over which the effective rest frame of matter is fitted. The perceived anisotropy therefore depends significantly on the redshift range and distribution of sources.
We find that the multipoles of the expansion rate, as well as the observer's velocity with respect to the large-scale matter flow, can be determined robustly with our approach. 
\end{abstract}

\keywords{Cosmology, Local Universe, $N$-body Simulations\\}

\section{Introduction}
\label{sec:intro} 

In modern cosmology, data are often analysed within a particular field theory and with 
assumptions about the energy-momentum content, for instance general relativity with ordinary matter, cold dark matter, radiation, and dark energy. A complementary way to analyse data is through cosmography,   
where the field theory of gravity is left unspecified. Cosmography thus presents an independent approach to constraining the kinematics and the curvature of the Universe, without relying on a particular dynamical theory \citep{Weinberg:1972kfs,1985PhR...124..315E}. 

Cosmographic approaches are particularly suitable for low-redshift measurements, but can also be applied in the intermediate- to high-redshift regime \citep[see e.g.][]{Cattoen:2007sk,Vitagliano:2009et,Capozziello:2020ctn}. 
The low-redshift regime of cosmology is sensitive to the large-scale structures within our cosmic neighbourhood. The anisotropies that are expected in cosmological observables due to cosmic structures are commonly modelled
using peculiar velocity corrections as predicted by a $\Lambda$ cold dark matter ($\Lambda$CDM) cosmology. This procedure has the disadvantage that one must trust that the derived corrections indeed give a complete description of the anisotropies in the observations. An alternative approach is to incorporate the anisotropies into the cosmography itself, thereby fitting for the anisotropies instead of making a model prediction for them. 
Such cosmographic treatments can determine the recent expansion history of the Universe 
in a model-independent way \citep{1985PhR...124..315E,Clarkson:2011br}. 
Moreover, estimates of the impact of local structure on the measurements on various scales might provide insights into cosmological tensions and anomalies seen across a growing number complementary cosmological data sets \citep{Perivolaropoulos:2021jda,Abdalla:2022yfr}.    

To formulate a cosmography that is valid for describing the anisotropy in our cosmological measurements, we must go beyond the spatially maximally-symmetric Friedmann-Lema\^itre-Robertson-Walker (FLRW) spacetimes and formulate observables geometrically in the absence of metric symmetries. 
Such formulations were pioneered by \citet{Kristian:1965sz} and later developed by, e.g.,  \citet{1970CMaPh..19...31M,Clarkson:2011uk,Clarkson:2011br,Umeh:2013UCT,Heinesen:2020bej,Heinesen:2021qnl,Maartens:2023tib}.  
The application of such frameworks to the analysis of data has so far been very limited \citep[though see][]{Dhawan:2022lze,Cowell:2022ehf} due to the high quality of data and good sky coverage necessary to measure anisotropic features in the observables robustly. 
However, as data sets continue to grow, cosmographic analyses accounting for the imprint of large-scale structure in observations are now within reach. 

Cosmographic analyses can be done with a variety of astrophysical observables. In this paper, we focus on angular diameter distance and analyse the anisotropic distance--redshift data for observers within $\Lambda$CDM cosmology. 
We use the \textit{UNITY} simulations presented in \citet{Coates:2020jzw}, which were performed with the general-relativistic $N$-body code \texttt{gevolution} \citep{Adamek:2015eda}. Using relativistic ray tracing we 
calculate distances and redshifts over the full skies of synthetic observers within the simulations. 
We investigate the distance--redshift cosmography from \citet{Heinesen:2020bej} as inferred by the observers, presenting constraints on the dominant multipoles in generalised cosmological parameters for different redshift intervals \citep[see][for estimates of the expected dominant multipoles]{Heinesen:2021azp,Cowell:2022ehf}.

Our high-resolution simulations present a realistic scatter of data points around any possible notion of a large-scale matter congruence. Here, a matter congruence is defined as a set of timelike integral curves that provide a threading of the observed patch of spacetime. To make such a notion more precise, we run a simulation where small-scale perturbations are filtered out in the initial conditions, leading to a situation where a smooth matter congruence is uniquely defined. Additionally, in a separate set of simulations we compare our results from \texttt{gevolution} to those found in the large-scale numerical relativity fluid simulations presented in \citet{Macpherson:2021gbh}.

In Section~\ref{sec:formalism}, we introduce the cosmography of \citet{Heinesen:2020bej} and recast the Taylor series expansion into an arguably more suitable Pad\'e approximant, as well as discuss fluctuations with respect to this smooth congruence model. In Section~\ref{sec:sims} we discuss details of the \textit{UNITY} simulations and ray tracing procedure, and in Section~\ref{sec:constraints} we outline our analysis of the synthetic observations. We present our main results in Section~\ref{sec:results} and conclude in Section~\ref{sec:conclusion}.

\section{Formalism} \label{sec:formalism}

In this section, we reformulate the anisotropic cosmo\-graphy of astrophysical sources relative to an observer in an arbirtary spacetime without exact symmetries. Following \citet{Heinesen:2020bej}, we first consider a situation where overlapping streams of the sources (shell crossing) can be ignored at the scales of interest, such that the astrophysical sources constitute a non-singular congruence of world lines generated by a 4--velocity field $u^\mu$.  
In this treatment we first assume that the observer is comoving with the congruence of sources, and then generalise the result to arbitrary observer frames. 
Furthermore, we assume that the photons emitted from the sources and received at the observer are described in the geometric optics limit, and that photon number is conserved, such that Etherington's reciprocity theorem applies. 

\subsection{Taylor series cosmographic expansion}\label{subsec:TSeries}

When Etherington's reciprocity applies the luminosity distance to the emitting source can be written as $d_L = d_A (1+z)^{2}$ for a light beam in a general geometry, where $d_A$ is the angular diameter distance (also known as the area distance) to the emitting source and $z$ is the redshift.

Consider the case where $z$ is a parameter along each of the photon null rays, such that the function $z \mapsto d_L(z,\boldsymbol{e})$ is well defined for each direction $\boldsymbol{e}$ of incoming light as seen on the observer's sky. 
If this function is analytic, its Taylor series expansion 
\begin{equation}\label{eq:series}
    d_L(z,\boldsymbol{e}) =  d_L^{(1)}(\boldsymbol{e})\, z   + d_L^{(2)} (\boldsymbol{e})\,z^2 +  d_L^{(3)}(\boldsymbol{e}) \,z^3 + \mathcal{O}( z^4) \, 
\end{equation} 
approximates the luminosity distance in some cosmic neighbourhood of the observer. 
The coefficients of the series expansion can be represented as 
\begin{equation}
\begin{aligned}
\label{eq:dLexpand2}
d_L^{(1)}(\boldsymbol{e}) &= \frac{1}{\Eu_o (\boldsymbol{e}) } \, , \qquad d_L^{(2)}(\boldsymbol{e}) =   \frac{1 - \mathfrak{Q}_o(\boldsymbol{e}) }{2 \Eu_o(\boldsymbol{e})}  \ ,\\
d_L^{(3)}(\boldsymbol{e}) &=  \frac{- 1 +  3 \mathfrak{Q}_o^2(\boldsymbol{e}) + \mathfrak{Q}_o(\boldsymbol{e})    -  \mathfrak{J}_o(\boldsymbol{e})   + \mathfrak{R}_o(\boldsymbol{e}) }{ 6  \Eu_o(\boldsymbol{e})}     \, , 
\end{aligned}
\end{equation}  
where the effective cosmological parameters 
\begin{eqnarray}\label{eq:paramseff}
    \Eu(\boldsymbol{e}) &\equiv& - \frac{1}{E^2}     \frac{ {\rm d} E }{{\rm d} \lambda}   \, , \qquad 
    \mathfrak{Q}(\boldsymbol{e})  \equiv - 1 - \frac{1}{E} \frac{     \frac{ {\rm d} \Eu}{{\rm d} \lambda}    }{\Eu^2}   \, , \nonumber\\ 
    \mathfrak{R}(\boldsymbol{e}) &\equiv&  1 +  \mathfrak{Q}  - \frac{1}{2 E^2} \frac{k^{\mu}k^\nu R_{\mu \nu} }{\Eu^2}   \, , \nonumber\\
    \mathfrak{J}(\boldsymbol{e})  &\equiv&   \frac{1}{E^2} \frac{      \frac{  {\rm d^2} \Eu}{{\rm d} \lambda^2}    }{\Eu^3}  - 4  \mathfrak{Q}  - 3, 
\end{eqnarray} 
generalise the FLRW Hubble, deceleration, curvature, and jerk parameters, respectively, in the distance--redshift cosmography. 
Here, $k^\mu$ is the photon 4--momentum, $\lambda$ is an affine parameter of the geodesic defined by $k^\mu \nabla_\mu \lambda = 1$, and the operator $\frac{ {\rm d}  }{{\rm d} \lambda} \equiv k^\mu \nabla_\mu$ is the directional derivative along the incoming null ray, $E = -u^\mu k_\mu$ is the photon energy as measured by the sources/observer, and $R_{\mu \nu}$ is the Ricci curvature of the spacetime.  
The effective cosmological Hubble, deceleration, jerk, and curvature parameters $\{\Eu,\mathfrak{Q},\mathfrak{J},\mathfrak{R}\}$ contain information about local kinematics and curvature effects that impact the luminosity distance--redshift relation and might for instance include anisotropic expansion of space, bulk flow of sources and lensing of photons. 

In this framework, the anisotropic
Hubble parameter can be written as the \textit{exact} multipole decomposition 
\begin{equation}\label{eq:Eevolution}
    \Eu(\boldsymbol{e}) = \frac{1}{3}\theta  - e^\mu a_\mu + e^\mu e^\nu \sigma_{\mu \nu}   \,  , 
\end{equation} 
where $\theta$ is the volume expansion rate in the source frame, $\sigma_{\mu \nu}$ is the anisotropic deformation rate of space as traced by matter (shear), $a^\mu \equiv u^\nu \nabla_\nu u^\mu$ is the 4--acceleration of the source frame, and $e^\mu$ denotes the spatial unit four-vector in direction $\boldsymbol{e}$ in that frame. 
The effective deceleration parameter can similarly be written in terms of the exact multipole decomposition 
\begin{multline}\label{eq:Q}
\mathfrak{Q}(\boldsymbol{e} ) = - 1 -  \frac{1}{\Eu^2(\boldsymbol{e} )} \biggl(
\overset{0}{\mathfrak{q}}   +  e^\mu  \overset{1}{\mathfrak{q}}_\mu   +    e^\mu e^\nu   \overset{2}{\mathfrak{q}}_{\mu \nu} \biggl.    \\
\biggr. +    e^\mu e^\nu e^\kappa \overset{3}{\mathfrak{q}}_{\mu \nu \kappa}   +   e^\mu e^\nu e^\kappa e^\lambda  \overset{4}{\mathfrak{q}}_{\mu \nu \kappa \lambda} \biggr)     \, , 
\end{multline}
with coefficients 
\begin{eqnarray*}
&& \overset{0}{\mathfrak{q}} \equiv  \frac{1}{3}   \frac{ {\rm d}  \theta}{{\rm d} \tau} + \frac{1}{3} D_{   \mu} a^{\mu  } - \frac{2}{3}a^{\mu} a_{\mu}    - \frac{2}{5} \sigma_{\mu \nu} \sigma^{\mu \nu}    \, ,  \nonumber\\ 
&& \overset{1}{\mathfrak{q}}_\mu \equiv  - \frac{1}{3} D_{\mu} \theta   -  \frac{2}{5}   D_{  \nu} \sigma^{\nu }_{\;  \mu  }   -  \frac{ {\rm d}  a_\mu }{{\rm d} \tau}  + a^\nu \omega_{\mu \nu}  +  \frac{9}{5}  a^\nu \sigma_{\mu \nu}     \, , \quad\nonumber\\
&& \overset{2}{\mathfrak{q}}_{\mu \nu}  \equiv     \frac{ {\rm d}  \sigma_{\mu \nu}   }{{\rm d} \tau} +   D_{  \la \mu} a_{\nu \ra }\! + a_{\la \mu}a_{\nu \ra } \! - 2 \sigma_{\kappa (  \mu} \omega^\kappa_{\nu )} \!  - \frac{6}{7} \sigma_{\kappa \la \mu} \sigma^\kappa_{\nu \ra }   \, ,\nonumber\\
\end{eqnarray*}
\begin{eqnarray}
\label{qpoles} 
 && \overset{3}{\mathfrak{q}}_{\mu \nu \kappa}  \equiv -  D_{ \la \mu} \sigma_{\nu   \kappa \ra }    -  3  a_{ \la \mu} \sigma_{\nu \kappa \ra }    \, , \qquad\qquad\qquad\qquad\nonumber\\
 && \overset{4}{\mathfrak{q}}_{\mu \nu \kappa \lambda}  \equiv  2   \sigma_{\la \mu \nu } \sigma_{\kappa \lambda \ra} \, , 
\end{eqnarray} 
where indices between angled brackets $\la \ra$
denote the traceless and symmetric part of the tensor in the involved indices, and $D_\mu$ is the spatial covariant derivative in the frame of the congruence. The operator $\frac{ {\rm d} }{{\rm d} \tau} \equiv u^\mu \nabla_\mu$ is the derivative along the world lines of the source congruence, and $\omega_{\mu \nu}$ is the vorticity tensor describing the rotation of the congruence. The higher-order effective cosmological parameters, $\mathfrak{J}$ and $\mathfrak{R}$, can similarly be written in terms of multipole expansions \citep[see][for their detailed expressions]{Heinesen:2020bej}.  

The radius of convergence of  Eq.~\eqref{eq:series} and the magnitude of the remainder for the truncation of the cosmography at third order depend on the particularities of the spacetime and the astrophysical survey geometry, and must in principle be assessed for each physical scenario of interest \citep[see, e.g.,][for estimates of the remainder terms within some numerical examples]{Macpherson:2021gbh}.

\subsection{Pad\'e approximant cosmography} \label{subsec:Pade}

We may take Eq.~\eqref{eq:series} as a first ansatz for an approximation of the cosmic neighbourhood of the observer. However, in the real Universe, the congruence described by our time-like vector field $u^\mu$ can be understood as capturing the coarse-grained dynamics at large scales. 
Individual astrophysical sources would then have peculiar motion with respect to this large-scale average, which would manifest as a scatter in the Hubble diagram.
The observed redshift therefore has particularly large random fluctuations which renders it discontinuous and non-monotonic along the lines of sight in the late-time, nonlinear Universe. On the other hand, a continuous and monotonic observable --- at least at low cosmological distances --- is the angular diameter distance. 
When considered as a function of the affine parameter along the line of sight, $d_A$ is mainly perturbed by weak gravitational lensing and by the relativistic aberration due to the peculiar motion of the observer. These perturbations are very smooth as long as no caustics are encountered on the light cone. In attempt to arrive at a framework which is useful for observational constraints in the late-time Universe, we therefore reformulate the distance--redshift relation using $d_A$ as the parameter in which the series expansion is performed. The inverse series of Eq.~\eqref{eq:series} reads
\begin{widetext}
\begin{eqnarray}
    z(\boldsymbol{e}, d_A) &=& \frac{1}{d_L^{(1)}\!(\boldsymbol{e})} d_A + \frac{2 d_L^{(1)}\!(\boldsymbol{e}) - d_L^{(2)}\!(\boldsymbol{e})}{\left[d_L^{(1)}\!(\boldsymbol{e})\right]^3} d_A^2 + \frac{5 \left[d_L^{(1)}\!(\boldsymbol{e})\right]^2\!\! + 2 \left[d_L^{(2)}\!(\boldsymbol{e})\right]^2\!\! - 6 d_L^{(1)}\!(\boldsymbol{e})\,d_L^{(2)}\!(\boldsymbol{e}) - d_L^{(1)}\!(\boldsymbol{e})\,d_L^{(3)}\!(\boldsymbol{e})}{\left[d_L^{(1)}\!(\boldsymbol{e})\right]^5} d_A^3+ \mathcal{O}(d_A^4)\,,\nonumber\\
    &=& \Eu_o(\boldsymbol{e}) d_A + \frac{1}{2} \left[\mathfrak{Q}_o(\boldsymbol{e}) + 3\right] \left[\Eu_o(\boldsymbol{e}) d_A\right]^2 + \frac{1}{6} \left[16 + 11 \mathfrak{Q}_o(\boldsymbol{e}) + \mathfrak{J}_o(\boldsymbol{e}) - \mathfrak{R}_o(\boldsymbol{e})\right] \left[\Eu_o(\boldsymbol{e}) d_A\right]^3 + \mathcal{O}(d_A^4)\,.
\label{eq:inverseseries}    
\end{eqnarray}
\end{widetext}

As suggested by \citet{Heinesen:2020bej}, it is convenient to define the following: 

\begin{subequations}\label{eqs:QJhat}
\begin{eqnarray}
    \hat{\mathfrak{Q}}_o(\boldsymbol{e}) &\equiv& \Eu^2_o(\boldsymbol{e}) \left[\mathfrak{Q}_o(\boldsymbol{e}) + 1\right]\,,\\
    \hat{\mathfrak{J}}_o(\boldsymbol{e}) &\equiv& \Eu^3_o(\boldsymbol{e}) \left[\mathfrak{J}_o(\boldsymbol{e}) - \mathfrak{R}_o(\boldsymbol{e}) - 1\right]\,,
\end{eqnarray}
\end{subequations}
which have a naturally-truncating, finite multipole decomposition within the cosmographic framework. 
Each multipole of these parameters contains physical information about the kinematical properties of the congruence $u^\mu$. For the case of $\hat{\mathfrak{Q}}_o(\boldsymbol{e})$, this can be seen in Eq.~\eqref{qpoles} \citep[see also the discussion in][]{Umeh:2013UCT}, while for $\hat{\mathfrak{J}}_o(\boldsymbol{e})$ this is given in Appendix~B of \citet{Heinesen:2020bej}. With the new quantities in Eqs.~\eqref{eqs:QJhat}, Eq.~\eqref{eq:inverseseries} simplifies to
\begin{multline}\label{eq:zTaylor}
    z(\boldsymbol{e}, d_A) = \Eu_o(\boldsymbol{e}) d_A + \left[\Eu_o^2(\boldsymbol{e}) + \frac{1}{2} \hat{\mathfrak{Q}}_o(\boldsymbol{e})\right] d_A^2 \\+ \left[\Eu_o^3(\boldsymbol{e}) + \frac{\hat{\mathfrak{J}}_o(\boldsymbol{e}) + 11 \Eu_o(\boldsymbol{e}) \hat{\mathfrak{Q}}_o(\boldsymbol{e})}{6}\right] d_A^3 + \mathcal{O}(d_A^4)\,.
\end{multline}

The third-order Taylor series expansions discussed in this section are expected to provide good approximations to the cosmographic model at very small distances. However, at larger distances the truncation error can increase very rapidly. An alternative series expansion is the Pad\'e approximant, which can often provide improved behaviour over a finite distance interval. Pad\'e approximants have already been proposed and implemented in the place of Taylor series expansions for more accurate constraints from distance--redshift data \citep[e.g.][]{Nesseris:2013bia,Sapone:2014nna,Demianski:2016dsa}, and have been shown to have radii of convergence beyond $z=1$ \citep{Gruber:2013wua}, which is the strict upper limit for the radius of convergence in Taylor series expansions.

\begin{figure}
    \includegraphics[width=\columnwidth, trim = 0mm 0mm 2mm 2mm, clip]{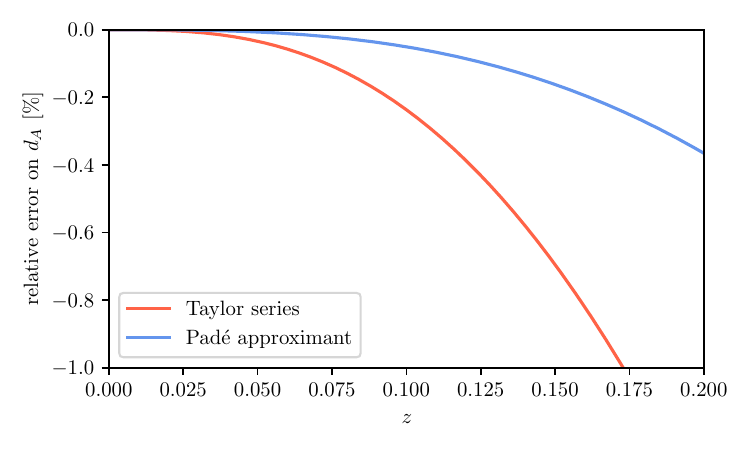}
    \caption{We show the relative error on the distance--redshift relation as a function of redshift for the third-order Taylor series expansion~\eqref{eq:zTaylor} and the $[3/2]$-order Pad\'e approximant~\eqref{eq:zPade} in red and blue, respectively. Both approximations use the same expansion rate, deceleration parameter and jerk, and were compared to an exact $\Lambda$CDM background solution with $\Omega_\mathrm{m} = 0.31$, $\Omega_\mathrm{rad} = 10^{-4}$, and vanishing curvature.}
    \label{fig:TaylorPade}
    \smallskip
\end{figure}

We propose the approximant
\begin{equation} \label{eq:zPade}
    z(\boldsymbol{e}, d_A) = \frac{\Eu_o(\boldsymbol{e}) d_A + \frac{1}{2} \hat{\mathfrak{Q}}_o(\boldsymbol{e}) d_A^2 + \frac{1}{6} \hat{\mathfrak{J}}_o(\boldsymbol{e}) d_A^3 + \ldots}{1 - \Eu_o(\boldsymbol{e}) d_A - \frac{4}{3} \hat{\mathfrak{Q}}_o(\boldsymbol{e}) d_A^2 - \ldots}\,,
\end{equation}
which is order $[3/2]$ and has the same Taylor series expansion as Eq.~\eqref{eq:zTaylor} up to $\mathcal{O}(d_A^4)$. Since we only fix the first three terms in the Taylor series, other Pad\'e approximants of order $[3/2]$ are possible, and our choice is mainly guided by the simplicity of the expression in terms of our preferred variables. 
For the FLRW background solution, the truncated Taylor series~\eqref{eq:zTaylor} has an error of $\approx 0.8\%$
with respect to the exact $\Lambda$CDM angular diameter distance at redshift $z\approx0.15$. Our expression~\eqref{eq:zPade} using a Pad\'e approximant with the same number of parameters reduces this error to better than $0.2\%$ (see Figure~\ref{fig:TaylorPade}).

\subsection{Fluctuations about the congruence} \label{subsec:dz}

Eq.~\eqref{eq:zPade} can be understood as a model for the `cosmographic' redshift of the underlying coarse-grained matter congruence. The observed redshift of astrophysical sources is {more realistically} given by the model plus some stochastic component $\delta z$, {representing the peculiar velocities of those sources with respect to the inhomogeneous congruence}, i.e.
\begin{equation}
    z_\mathrm{obs} = z(\boldsymbol{e}, d_A) + \delta z\,.
\end{equation}
This expression {combined with Eq.~\eqref{eq:zPade}} generalises the common interpretation of observed redshifts in an FLRW cosmology to the inhomogeneous case.

Here we are still assuming that the observation is made in the rest frame of the congruence. 
In reality, the observer can have some velocity with respect to the large-scale congruence which may even need to be determined from the same data. It is therefore more useful to describe the observation in an arbitrary frame and allow for three additional parameters that characterise the boost between the frames. 
These three parameters are included in our joint parameter fit, and thus no knowledge about the velocity of the observer is assumed.

We denote the peculiar three-velocity vector of the observer with respect to the congruence $u^\mu$ as $\boldsymbol{\beta}$, with length $\beta = v/c$. Special relativity implies that the observed redshift in the boosted frame, $z'_\mathrm{obs}$, is given by
\begin{equation}
    1 + z'_\mathrm{obs} = \frac{1 - \boldsymbol{e}'\cdot\boldsymbol{\beta}}{\sqrt{1-\beta^2}} \left(1+z_\mathrm{obs}\right)\,,
\end{equation}
where $\boldsymbol{e}'$ is the observed position in the boosted frame. The latter is different from $\boldsymbol{e}$ in the rest frame of the congruence due to relativistic aberration. They are related by
\begin{equation}
    \boldsymbol{e} = \frac{\sqrt{1-\beta^2}}{1-\boldsymbol{e}'\cdot\boldsymbol{\beta}} \boldsymbol{e}' - \frac{1 - \left(1 - \sqrt{1-\beta^2}\right) \beta^{-2} \boldsymbol{e}'\cdot\boldsymbol{\beta}}{1-\boldsymbol{e}'\cdot\boldsymbol{\beta}} \boldsymbol{\beta}\,.
\end{equation}
The aberration also changes the angular distance which follows from $d_A^2 d\Omega = \left(d'_A\right)^2 d\Omega'$, where $d\Omega$ is the solid angle element. We therefore find the well-known result that
\begin{equation}
    d_A = \frac{1-\boldsymbol{e}'\cdot\boldsymbol{\beta}}{\sqrt{1-\beta^2}} d'_A\,.
\end{equation}
Putting everything together we have a model for the observed redshift in the boosted frame. If the velocity is small, i.e.\ $v \ll c$, we find
\begin{multline}
\label{eq:boostexpansion}
    z'_\mathrm{obs} \approx \left(1-\boldsymbol{e}'\cdot\boldsymbol{\beta}\right) z\left[\left(1+\boldsymbol{e}'\cdot\boldsymbol{\beta}\right) \boldsymbol{e}' - \boldsymbol{\beta}, \left(1-\boldsymbol{e}'\cdot\boldsymbol{\beta}\right) d'_A\right]\\-\boldsymbol{e}'\cdot\boldsymbol{\beta} + \left(1 - \boldsymbol{e}'\cdot\boldsymbol{\beta}\right) \delta z\,.
\end{multline} 
The aberration of the original anisotropy pattern is a second-order effect since we may write 
\begin{equation}
z(\boldsymbol{e}, d_A) = z(\boldsymbol{e}', d_A) + (\boldsymbol{e} - \boldsymbol{e}') \cdot \boldsymbol{\nabla}_{\boldsymbol{e}'} z(\boldsymbol{e}', d_A)  + \ldots\,,
\end{equation}
with $\boldsymbol{e} - \boldsymbol{e}' \approx (\boldsymbol{e}'\cdot\boldsymbol{\beta}) \boldsymbol{e}' - \boldsymbol{\beta}$. The second- and higher-order corrections in above equation are therefore strongly suppressed for the low multipoles of the distance--redshift relation we are interested in, allowing us to make the further approximation that $z(\boldsymbol{e}, d_A) \approx z(\boldsymbol{e}', d_A)$. 
If we also drop the term $\beta\times\delta z$ from Eq.~\eqref{eq:boostexpansion}, which is quadratic in velocities, we finally arrive at
\begin{equation}
\label{eq:zmodel_final}
    z'_\mathrm{obs} \approx \left(1-\boldsymbol{e}'\cdot\boldsymbol{\beta}\right) z\left[\boldsymbol{e}', \left(1-\boldsymbol{e}'\cdot\boldsymbol{\beta}\right) d'_A\right] -\boldsymbol{e}'\cdot\boldsymbol{\beta} +  \delta z\,.
\end{equation}

This is the expression we use to constrain the anisotropies in our cosmography model \eqref{eq:zPade} using the observed redshifts, angular diameter distances, and sky positions measured by the observers 
in the simulation. We describe the process of constraining the parameters of the cosmography in Section~\ref{subsec:MCMC}.

\section{Simulations} \label{sec:sims}

In order to assess how well the parameters encoding the anisotropy of the distance--redshift relation can be measured when assuming the cosmographic model presented in the previous section, we use synthetic source catalogues that we generate from simulations. In Section~\ref{subsec:unity}, we describe the existing \textit{UNITY} simulation data we use for our main analysis. In Section~\ref{subsec:smoothUNITY} we introduce new, low-resolution simulations, which represent `smooth' analogues of the \textit{UNITY} simulations from our main analysis.

\subsection{\textit{UNITY} simulation suite} \label{subsec:unity}

We use data from existing simulations described in \citet{Coates:2020jzw}: the \textit{UNITY} simulation suite consists of 34 independent realisations of a flat $\Lambda$CDM Universe with cosmological parameters $h = 0.67$, $\Omega_\mathrm{b} = 0.049$, $\Omega_\mathrm{c} = 0.27$, $T_\mathrm{CMB} = 2.7255\,\mathrm{K}$ and a minimal-mass normal hierarchy of neutrinos with $\sum m_\nu = 0.06\,\mathrm{eV}$. Gaussian initial conditions follow a power law for primordial perturbations with {amplitude} $A_s = 2.1\times 10^{-9}$ at the pivot scale of $0.05\,\mathrm{Mpc}^{-1}$ and spectral index $n_s = 0.96$. The simulated volume is a cube of $4032\,\mathrm{Mpc}/h$ on each side, and the matter (baryons and cold dark matter are treated as a single species) is sampled using $2304^3$ mass elements that form a collisionless $N$-body ensemble. In the present analysis we only use one of the realisations available, but the remaining ones could be analysed in the future to generate an ensemble of observers and quantify the cosmic variance in the cosmographic parameters.

The simulations were performed using the $N$-body code \texttt{gevolution} \citep{Adamek:2015eda} which adopts a weak-field expansion of general relativity to incorporate relativistic effects. The numerical output consists of data for the matter distribution and metric perturbations on the past light cone of an observer placed at a random location inside the simulation. In the present study we therefore do not consider any specific observer selection effects: as our synthetic observations span hundreds of megaparsecs, we expect such selection effects to be rather unimportant. We ran some quick checks on a few other realisations to convince ourselves that the realisation we use here is not a statistical outlier.

For convenience, the data are discretised in terms of concentric shells around the observer that are evenly spaced in comoving coordinates at intervals of $1.75\,\mathrm{Mpc}/h$. Each shell is pixelised using the \texttt{HEALPix} framework \citep{Gorski:2004by} at a resolution adapted to the surface area of the shell in order to sample the simulated matter and metric perturbations as close as possible to the resolution of $1.75\,\mathrm{Mpc}/h$, which corresponds to the Cartesian mesh used in the particle-mesh algorithm of the code.

The discretised data are post processed using a simple ray tracer that works in the Born approximation \citep[see][for details of the ray-tracing procedure]{Lepori:2020ifz}. Assuming that the lines of sight are unperturbed has the benefit that observables can be computed very efficiently by essentially performing weighted sums of pixel maps corresponding to consecutive shells down the past light cone. The ray tracer computes two observables for each pixel: the observed redshift $z_\mathrm{obs}'$ and the angular distance $d_A'$, where here the primes indicate that the observation is done in the rest frame corresponding to the time slicing of Poisson gauge, which closely matches the rest frame of the cosmic microwave background. In a Universe with inhomogeneous matter, this frame does in general not coincide with the local rest frame of the matter flow, i.e.\ one expects to see some bulk flows with respect to the cosmic microwave background. In our model, this is accounted for by the observer boost $\boldsymbol{\beta}$.

Each observable includes all relativistic perturbations at first order. For the observed redshift these are the Doppler shift due to peculiar motion of the source, the gravitational redshift, the integrated Sachs-Wolfe effect and the Shapiro time delay, while the angular distance is only perturbed by the weak-lensing convergence. In order to assign the peculiar velocity we take a mass-weighted average of the velocity of 
all particles in each pixel. It is worth pointing out that, although all effects are added linearly by the ray tracer, they arise from a completely non-perturbative underlying solution.

For this work we assume that astrophysical sources are good tracers of the underlying matter, i.e.\ we sidestep the issue of bias and use a density of sources that is directly proportional to the matter density. This also avoids the issue of empty pixels in which no unique velocity can be assigned.

\subsection{`Smooth' simulations} \label{subsec:smoothUNITY}

The simulations described in the previous section contain small-scale nonlinear structures at low redshift, thus, constructing a large-scale congruence description of these simulations is non-trivial \citep[e.g.][]{Clarkson:2011zq}. We attempt to do this by directly fitting the cosmographic framework outlined in Section~\ref{sec:formalism}, which essentially introduces a `smoothing length' corresponding to the redshift range of the data we include. 
We also perform the same fits using a simulation which \textit{should} have a uniquely defined
large-scale congruence description. By comparing this fit to direct calculations of the effective cosmological parameters \eqref{eq:paramseff} in the same simulation \citep[in a manner equivalent to those presented in][]{Macpherson:2021gbh}, we can assess how well the fit is matching the `true' congruence description of the simulation. 

We take the \textit{UNITY} simulation outlined in the previous section and remove all perturbations with wavelengths shorter than $670\,\mathrm{Mpc}$ from the initial data by applying a low-pass filter. We decrease
the spatial resolution of this simulation by a factor of four and the mass resolution by a factor of eight ---
justified by the lack of small-scale power in the simulation. Aside from this resolution change, we perform the simulation with parameters as described above. 
The filter is applied only to the initial data, and naturally some structures beneath this scale will form through the course of the simulation. However, the power on small scales will be significantly reduced at late times with respect to the original \textit{UNITY} simulations. We thus consider the scale of $670\,\mathrm{Mpc}$  
as an approximate smoothing scale for the late-time simulation, which is much larger than the resolution scale of approximately $10\,\mathrm{Mpc}$. For this simulation, we use the analysis code of \citet{Macpherson:2021gbh} to 
calculate the effective cosmological parameters \eqref{eq:paramseff} by taking numerical derivatives of dynamical quantities at the observer position. We then use the data obtained by ray tracing 
to constrain these same parameters and compare the results. 

\citet{Macpherson:2021gbh} performed their analysis on simulations from the \texttt{Einstein Toolkit} \citep{Loffler:2011ay}. To check for consistency when changing the simulation framework, we replicated one of those simulations with \texttt{gevolution} and repeated the analysis. The \texttt{Einstein Toolkit} uses a fluid approximation for the matter field in numerical relativity while 
\texttt{gevolution} uses an $N$-body particle description in weak-field general relativity. We find good agreement between these two cases (see Appendix~\ref{appx:ETcomp}) and thus conclude that, under the right conditions, we can find a large-scale congruence description of the $N$-body particle simulation data. 

It is worth pointing out that due to the non-commutation of averaging and time evolution in nonlinear general relativity, we do not in general expect the large-scale congruence of the \textit{UNITY} simulation to coincide with that of the simulation with smoothed initial data.

\section{Parameter constraints} \label{sec:constraints}

In this section we discuss details of our method for constraining the anisotropic cosmography using synthetic observations from the \textit{UNITY} simulations. First, we describe the assumptions we make and how they reduce the degrees of freedom of the model cosmography, and then we outline the pipeline we use to constrain these degrees of freedom using simulation data.

\subsection{Dominant multipoles} \label{subsec:multipoles}

The effective cosmological parameters in the Pad\'e approximant \eqref{eq:zPade} of the distance--redshift relation contain a hierarchy of multipoles. The multipole decompositions of the Hubble and deceleration parameters are given in \eqref{eq:Eevolution} and \eqref{eq:Q}, respectively, and the curvature and jerk parameter decompositions can be found in \citet{Heinesen:2020bej}. This full hierarchy contains many degrees of freedom \citep[though see][for a reduction in the degrees of freedom by a factor of two]{Heinesen:2021azp}, and constraining all of  them with current data is difficult. We can make some simplifying assumptions to reduce the number of degrees of freedom and allow the best possible constraints. For a
congruence of collisionless matter the acceleration vector $a^\mu$ vanishes, which removes the dipole component of the anisotropic Hubble parameter\footnote{Figure~\ref{fig:2Dplots} in Appendix~\ref{appx:ETcomp} shows that the contribution of the acceleration is far subdominant to the shear and expansion.}.
In this work, we thus focus on constraining the monopole and quadrupole of the Hubble parameter as well as the independent multipoles of the deceleration parameter (monopole, dipole, quadrupole and octopole). According to Eq.~\eqref{qpoles}, the hexadecapole of the deceleration parameter is fully determined by the quadrupole of the Hubble parameter, and we build this constraint into our model. We fix $\hat{\mathfrak{J}}_o = 0$ which is close to the value for a flat FLRW universe with a realistic value of the present energy density of radiation, $\hat{\mathfrak{J}}_o/H_0^3 =  2\,\Omega_\mathrm{rad} \approx 10^{-4}$. Note the ``$-1$'' that enters the definition of $\hat{\mathfrak{J}}_o$ in Eq.~\eqref{eqs:QJhat} to see the connection to the canonical jerk parameter.

\subsection{Markov chain Monte Carlo (MCMC) analysis} \label{subsec:MCMC}

We constrain the components of each anisotropy tensor by modelling the logarithm of the likelihood as
\begin{equation}
    \ln\mathcal{L} = -\frac{1}{2}\sum_i \left[\frac{\delta z^{(i)}}{\sigma\left(d_A^{(i)}\right)}\right]^2 - \sum_i \ln \sigma\left(d_A^{(i)}\right)\,,
\end{equation}
up to an irrelevant constant. In this formula, we compute the residual $\delta z^{(i)}$ for each individual source from Eq.~\eqref{eq:zmodel_final}, i.e.,\ we solve that equation for $\delta z$ using the observed redshift $z'_\mathrm{obs}$, the observed direction $\boldsymbol{e}'$ and the observed angular distance $d_A'$ of each observed pixel $i$. The model parameters are given by the boost velocity vector $\boldsymbol{\beta}$ (three components) and all the parameters that enter Eq.~\eqref{eq:zPade} describing the cosmography (22 independent parameters after fixing the four-acceleration and the jerk). Furthermore, we add two nuisance parameters to model the Gaussian scatter $\sigma(d_A)$,
\begin{equation}\label{eq:scatter}
    \sigma\left(d_A\right) = \frac{\sigma_0}{1 + s_1 d_A} \approx \frac{\sigma_0}{1 + s_1 \left(1-\boldsymbol{e}'\cdot\boldsymbol{\beta}\right) d'_A}\,,
\end{equation}
in order to allow a certain evolution of the scatter with distance. This means that the Gaussian scatter needs to be evaluated for each pixel separately, and we try to find a global fit for $\sigma_0$ and $s_1$. The underlying assumption of this likelihood model is that $\delta z$ is indeed a Gaussian random variable with zero expectation value and a variance that mildly depends on angular distance. Our results show that this is a reasonable assumption in the presence of small-scale nonlinear structure, but not such a good approximation for the case where we filtered the initial conditions. Details are discussed in the next section.

Our data vectors consist of randomly selected pixels sampling the full sky at redshifts below a chosen threshold.
We set the probability for selecting a pixel proportional to its comoving matter density. Since we want to model the distance--redshift relation over a certain baseline distance, we want the sources roughly evenly distributed across that distance. The volume factor naturally favours objects at large distance, and we compensate this fact by scaling the probability by an additional factor proportional to the solid angle of the pixel. In this way, for any given solid angle of the observations, the data are roughly evenly distributed in distance. This is of course just one possible selection function, and we expect that different selection functions may lead to different results depending on how much they favour low-redshift sources over high-redshift ones, or vice versa. We use different redshift thresholds to gain some first insights into this effect.

Our analysis does not require a lower redshift cutoff and we even include sources with negative redshift. In fact, it may be important to keep such sources for an accurate fit of the observer velocity.
We vary the redshift range of sources in our synthetic catalogs up to a maximum of $z=0.15$ to ensure a small truncation error in our Pad\'e approximant. Keeping the number of sources at low redshift approximately constant, 
this results in a different number of sources as we increase the maximum redshift.
We tested different values for the density of sources, and as one expects, a larger number of sources generally leads to tighter posterior contours while remaining consistent with one another. Our final choice strikes a balance between constraining power and cost of running the MCMC chains, while also being guided by the typical size of existing source catalogs from real observations. In the \textit{UNITY} simulation the number we used ranges from about $2000$ sources at $z=0.04$ up to about $7500$ sources at $z = 0.15$. For the `smooth' simulation we have fewer sources, ranging between about $500$ at $z = 0.025$ to about $2000$ at $z = 0.15$.

\section{Results} \label{sec:results}

\begin{figure}
    \includegraphics[width=\columnwidth, trim = 0mm 0mm 10mm 12mm, clip]{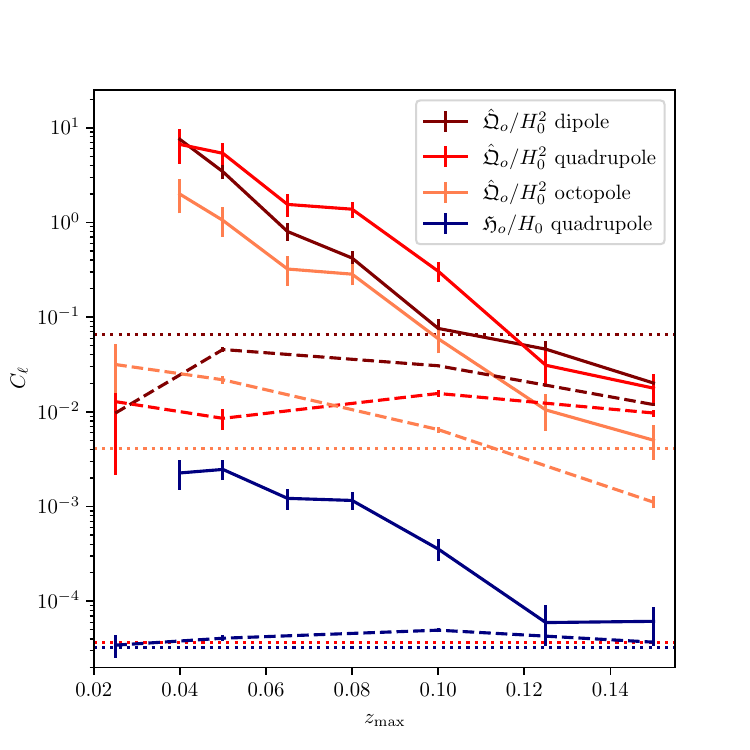}
    \caption{Multipole power $C_\ell = \left(2\ell + 1\right)^{-1} \sum_m \left\vert a_{\ell m}\right\vert^2$ for the quadrupole ($\ell = 2$) of $\Eu_o / H_0$ which measures the shear tensor of the coarse-grained matter congruence relative to the expansion, and the independent multipoles of $\hat{\mathfrak{Q}}_o / H_0^2$ that characterise the anisotropy in the deceleration parameter. The error bars indicate $95\%$ confidence intervals. For all anisotropy parameters the power drops nearly exponentially as the fitted range in redshift $z$ increases (horizontal axis). The dashed lines show results from a simulation where perturbations below the scale of 670\,Mpc were filtered out in the initial data, and hence a well-defined matter congruence exists. For this case, the horizontal dotted lines show an alternative measurement using the properties of null geodesics at the observer location ($z \rightarrow 0$).}
    \label{fig:Cls}
    \smallskip
\end{figure}
Figure~\ref{fig:Cls} shows the multipole power $C_\ell= \left(2\ell + 1\right)^{-1} \sum_m \left\vert a_{\ell m}\right\vert^2$ as a function of the maximum redshift $z_\mathrm{max}$ of the fitting interval. Solid curves represent constraints in the fully nonlinear \textit{UNITY} simulation and dashed curves represent constraints in the 
simulation with filtered initial data.
We show the power in the quadrupole in $\mathfrak{H}_o/H_0$ (dark blue) as well as the dipole (brown), quadrupole (red), and octopole (orange) in $\hat{\mathfrak{Q}}_o/H_0^2$, where $H_0 = \theta/3$ is the monopole of the expansion rate $\mathfrak{H}_o$. The error bars indicate $95\%$ confidence intervals. 
The smaller error bars in the `smooth' case are due to the much reduced scatter in the data, owing to the lack of small-scale perturbations. The horizontal dotted lines (without error bars) indicate the multipole amplitudes found by taking numerical derivatives along null rays at the observer location in the `smooth' simulation, following the analysis of \citet{Macpherson:2021gbh} (see Appendix \ref{appx:ETcomp} for more details). 

The `smooth' simulations have structure below 670\,Mpc removed from the initial data, corresponding to the diameter of the light cone at $z\approx 0.08$. Consequently there is relatively little evolution of the multipole amplitude across the entire redshift range probed. The opposite is true for the simulation that includes small-scale nonlinear structure: anisotropies are very large if the fitting range includes only very low redshifts, and they drop nearly exponentially as the fitting range is enlarged. However, the anisotropy levels always remain larger than in the case of the smoothed initial data, even beyond the smoothing scale. 
We reiterate that the initial data of these simulations are otherwise identical; removing structure beneath 670\,Mpc is the only difference.

The dotted lines in Figure~\ref{fig:Cls}, which are obtained by taking numerical derivatives of the distance--redshift relation at the observer location in the `smooth' simulation, are mostly in fair agreement with our MCMC analysis, with the notable exception of the quadrupole of $\hat{\mathfrak{Q}}_o$ where the measurements appear to disagree by more than two orders of magnitude.
In this particular case, the apparent disagreement is a result of projecting the posterior onto $C_\ell$, which is a quadratic quantity. We show in Figure~\ref{fig:quadrupoles} (discussed in more detail below) that the posterior of the quadrupole components of $\hat{\mathfrak{Q}}_o$ at low redshift encloses the zero quadrupole value within its $1\,\sigma$-contours. However, $C_\ell$ effectively measures the square distance from zero in the five-dimensional parameter space of quadrupole components. The steep increase in volume as one moves away from zero initially overcomes the decrease in likelihood, giving a large value for the marginalised $C_\ell$ but also an error bar of similar order. This effect is problematic for multipoles whose full posterior overlaps with the origin. On the other hand, if the entire posterior can be seen to be far away from zero, the $C_\ell$ become a useful measure of the distance.

The physical reason for the expected smallness of the quadrupole term relative to the dipole and octopole of $\hat{\mathfrak{Q}}_o$ can be seen in Eq.~\eqref{qpoles}: while the dipole and octopole are dominated by spatial gradients of the matter flow field, the leading term of the quadrupole is the time derivative of the shear (assuming a geodesic flow). For cosmological matter perturbations, time derivatives are generally much smaller than spatial gradients.

\citet{Cowell:2022ehf} placed an upper limit of $2.88\%$ on the amplitude of the quadrupole in the Hubble parameter --- relative to the monopole --- using Pantheon+ supernovae data. The data in this work lay in the redshift range $z=0.023$ to $z=0.15$. The amplitude for a maximum redshift of $z=0.15$ that we find from $C_2$ in Figure~\ref{fig:Cls} is 
$\lesssim 1.6\%$ (at $95\%$ confidence), which is consistent with the current upper limit from supernovae and suggests that a detection should be possible in the near future. We note this amplitude is similar for both the \textit{UNITY} and `smooth' simulations we use here.

\begin{figure}
    \includegraphics[width=\columnwidth]{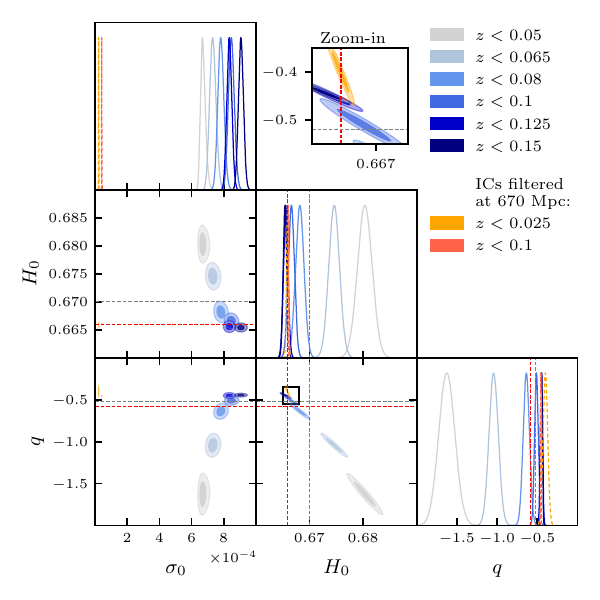}
    \caption{2D marginalised posteriors of $\sigma_0$, $H_0$ (in units of $100\,\mathrm{km}\,\mathrm{s}^{-1}\,\mathrm{Mpc}^{-1}$) and $q$ for different fitting ranges in redshift $z$ (different colours). The sampling is roughly uniform in observed distance $d_A$ to the sources. Results from a simulation where perturbations below the scale of 670\,Mpc were filtered out in the initial data are shown in orange and red. For this case, the red dashed lines show an alternative measurement of $H_0$ and $q$ using numerical derivatives along null geodesics at the observer location ($z \rightarrow 0$). The values for the fiducial FLRW background model are indicated by grey dashed lines. The inset labelled ``Zoom-in'' magnifies a region highlighted in the 2D plot for the parameters $H_0$ and $q$.}
    \label{fig:Hq}
    \smallskip
\end{figure}
Figure~\ref{fig:Hq} shows two-dimensional marginalised posterior distributions of the monopole of the Hubble parameter, $H_0$, and the monopole of the deceleration parameter, $q \equiv -1 -\overset{0}{\mathfrak{q}} /H_0^2$. The exact monopole of $\mathfrak{Q}_o$ would receive small corrections due to the anisotropy of $\Eu_o$, but we neglect these higher-order terms here. In other words, $q$ parameterises the exact monopole of $\hat{\mathfrak{Q}}_o$ which has a direct interpretation in terms of kinematical quantities of the cosmographic model as specified in Eq.~\eqref{qpoles}. Figure~\ref{fig:Hq} also includes posteriors for the nuisance parameter $\sigma_0$ 
defined in \eqref{eq:scatter}. Grey to progressively darker blue contours show an increasing maximum redshift of the sources as stated in the legend. Orange and red contours show constraints in the `smooth' simulations. Grey dashed lines in the panels with $H_0$ and $q$ represent the true values of these parameters in the FLRW limit of the simulations, while the red dashed lines indicate the corresponding measurements based on numerical derivatives at the observer location in the `smooth' simulation.

As a general trend, the posterior contours tend to move through parameter space and progressively shrink as the redshift range used in the fit increases.
While the tightening of contours is partly due to the increase of the number of sources, the path of the best fit illustrates the dependence on the sampled volume.
At low redshift, $q$ is particularly poorly constrained, as may be expected, but even at $z = 0.15$ the contours have not yet converged fully to the FLRW limit. For the `smooth' simulation, the contours at low and high redshift are roughly compatible, and both of them are in fair agreement with the \textit{UNITY} simulation when data up to $z = 0.1$ or higher are included in the fit, see the panel labelled ``Zoom-in'' in Figure~\ref{fig:Hq}. The fitted value for $\sigma_0$ for \textit{UNITY} is clustered around $8\times 10^{-4}$ which is commensurate with the expected velocity dispersion. In the case with filtered initial data the scatter is significantly smaller, of the order of few $\times 10^{-5}$.

The monopole values found via numerical derivatives (red dashed lines in Figure~\ref{fig:Hq}) should correspond closely to the constrained parameters in the `smooth' simulation case. While we find good agreement in the case of $H_0$, the deceleration parameter shows disagreement between these two calculations at several $\sigma$.

\begin{figure}
    \includegraphics[width=\columnwidth, trim = 0mm 0mm 0mm 12mm, clip]{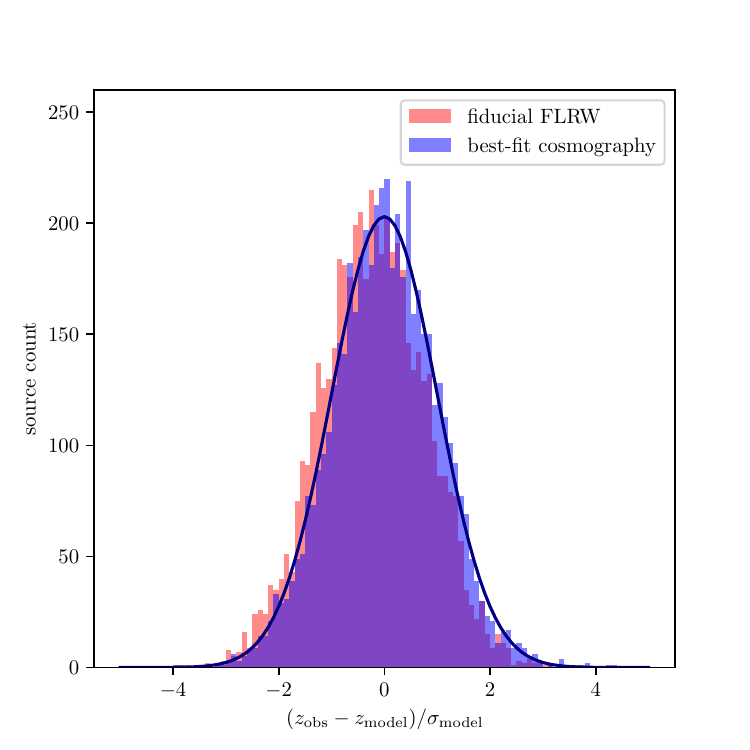}
    \caption{Scatter of individual sources around the modelled distance--redshift relation. The best-fit cosmography is shown in blue, and the scatter is normalised according to the best-fit Gaussian inferred in the likelihood analysis (dark blue line). The fiducial FLRW model is shown in red, and we normalise the scatter by the standard deviation in this case, i.e.\ without assuming a model. The histogram shows that the assumed Gaussian likelihood model provides a fair description of the data, here for the case of approximately 5000 sources with redshifts $z < 0.1$.}
    \label{fig:histscatter}
    \smallskip
\end{figure}

Figure~\ref{fig:histscatter} shows a histogram of the observed scatter in redshift of the individual sources in the \textit{UNITY} simulation with respect to the best-fit cosmographic model, Eq.~\eqref{eq:zPade}, and the fiducial FLRW model (blue and red, respectively).
In the former case, the scatter of each source is normalised to the inferred standard deviation $\sigma\left(d_A^{(i)}\right)$ of the best-fit model, Eq.~\eqref{eq:scatter}, whereas in the latter case we normalise the scatter to the standard deviation of the entire distribution. 
We find an excellent agreement with the Gaussian model assumed in the likelihood analysis (dark blue curve), and the normality test using the Kolmogorov-Smirnov statistic is passed easily with a p-value of $0.2$ (rising to $0.9$ if the sample mean is subtracted). The scatter with respect to the fiducial FLRW model is also Gaussian, yet with a standard deviation that is about $19\%$ larger than for the best-fit cosmography. In addition, the central value of the distribution is biased significantly --- the Kolmogorov-Smirnov test rules out a zero-centered Gaussian at more than $10\,\sigma$ in this case.

\begin{figure}
    \includegraphics[width=\columnwidth, trim = 0mm 0mm 0mm 12mm, clip]{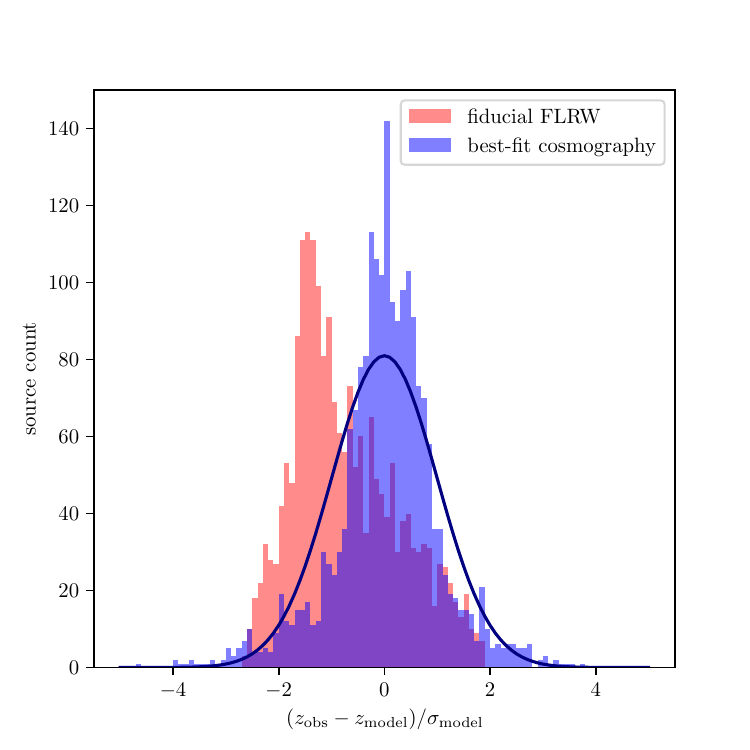}
    \caption{Same as Fig.~\ref{fig:histscatter}, but for the simulation with filtered initial conditions and using approximately 2000 sources with redshifts $z < 0.1$. The Gaussian likelihood model is clearly not a very good description of the data in this case.}
    \label{fig:smoothedhistscatter}
    \smallskip
\end{figure}

A corresponding histogram for the `smooth' simulation is shown in Figure~\ref{fig:smoothedhistscatter}. Since there is no intrinsic velocity dispersion in this model the small residual redshift errors are dominated by non-Gaussian numerical noise due to discretisation. We confirm this by conducting a resolution study, showing that the scatter almost doubles when we decrease the resolution of the `smooth' simulation by a factor of two. 
As a consequence, the Gaussian likelihood model is not a good approximation in this case, and we caution that the corresponding posteriors of our MCMC analysis should therefore be understood as being only indicative. While the mean of the data can still be inferred accurately, the model provides only a poor description of the scatter around that mean, making it hard to interpret the size of the posterior contours. 

It is interesting to note that the cosmographic model provides a better and less biased description of the data than the fiducial FLRW model in all cases shown in Figures~\ref{fig:histscatter} and \ref{fig:smoothedhistscatter}. We attribute this to the fact that the cosmographic model already accounts for some of the departure from the homogeneous and isotropic limit by fitting an inhomogeneous coarse-grained matter congruence to the observations. This provides a better physical model for the distribution of sources.

\begin{figure*}
    \centering
    \includegraphics[width=\textwidth]{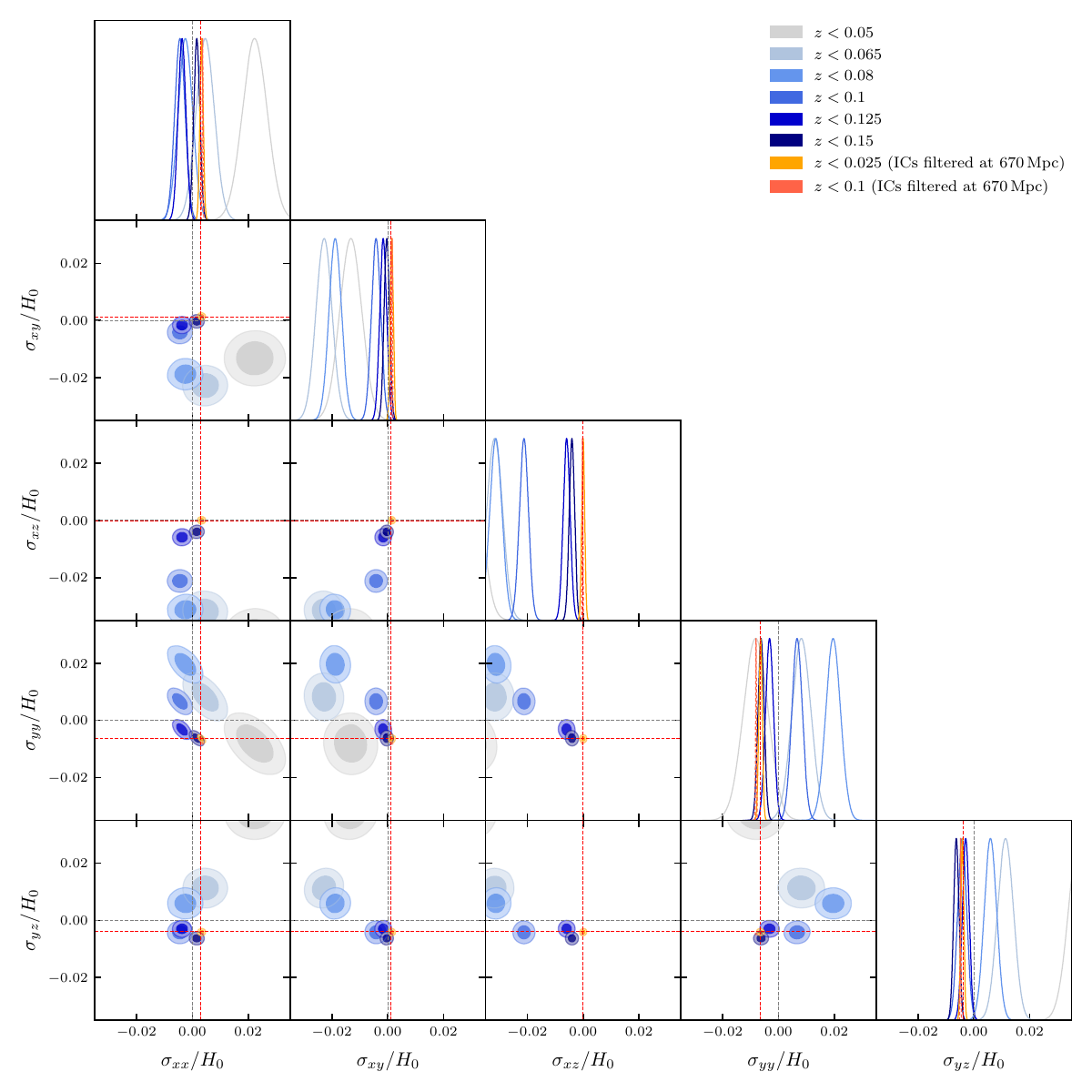}
    \caption{2D marginalised posteriors of the quadrupolar contributions to $\Eu_o$ for different fitting ranges in redshift $z$ (different colours). The sampling is roughly uniform in observed distance $d_A$ to the sources. The five independent components of $\sigma_{\mu\nu}$, which is the shear tensor of the fitted coarse-grained matter congruence, govern the quadrupole anisotropy of the Hubble parameter. Results from a simulation where perturbations below the scale of 670\,Mpc were filtered out in the initial data are shown in orange and red. For this case, the red dashed lines show an alternative measurement using the properties of null geodesics at the observer location ($z \rightarrow 0$). All shown quantities would be zero in an exact FLRW model (grey dashed lines).}
    \label{fig:shear}
\end{figure*}

Figure~\ref{fig:shear} shows two--dimensional posterior contours for the five independent components of the quadrupole of $\mathfrak{H}_o$, the shear tensor. The anisotropy in the expansion rate is typically of the order of a few percent in the redshift ranges studied, and systematically decreases towards larger redshift ranges. It is worth pointing out that the contours generally exclude the isotropic FLRW model (grey dashed lines) with high significance, and that the convergence towards isotropy does not follow a straight path in parameter space. This means that the best-fit quadrupole shape depends quite strongly on the redshift window over which it is inferred. The red dashed lines in Figure~\ref{fig:shear} indicate the quadrupole shape determined from numerical derivatives at the observer location within the `smooth' simulation. The corresponding contours from the MCMC analysis (very tight orange and red contours) are consistent with themselves over varying redshifts and agree extremely well with this alternative measurement. We remind the reader that the confidence contours should not be taken at face value in this case, since we know that the Gaussian likelihood model is not a good description.

\begin{figure*}
    \centering
    \includegraphics[width=\textwidth]{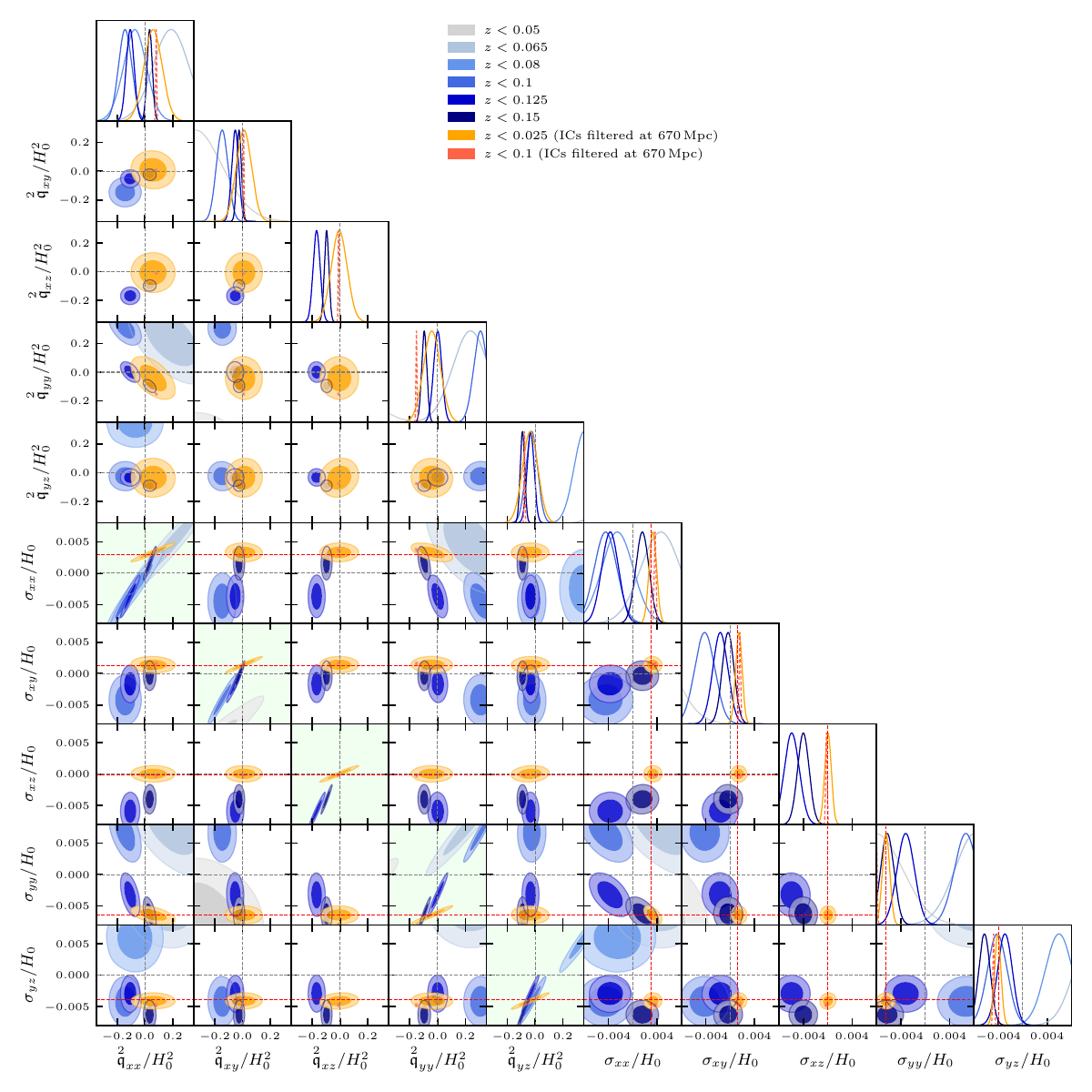}
    \caption{2D marginalised posteriors of the quadrupolar contributions to the distance--redshift relation for different fitting ranges in redshift $z$ (different colours). The sampling is roughly uniform in observed distance $d_A$ to the sources. As in Figure~\ref{fig:shear}, the five independent components of $\sigma_{\mu\nu}$ govern the quadrupole anisotropy of the Hubble parameter, whereas the five components of $\overset{2}{\mathfrak{q}}_{\mu\nu}$ characterise the quadrupole anisotropy of the deceleration parameter. Results from a simulation where perturbations below the scale of 670\,Mpc were filtered out in the initial data are shown in orange and red. For this case, the red dashed lines show an alternative measurement using the properties of null geodesics at the observer location ($z \rightarrow 0$). All shown quantities would be zero in an exact FLRW model (grey dashed lines). The panels highlighted in light green show the degeneracies between pairs of parameters that affect the same quadrupole anisotropy.}
    \label{fig:quadrupoles}
\end{figure*}
Figure~\ref{fig:quadrupoles} shows the same two-dimensional posteriors of the quadrupole components of $\mathfrak{H}_o$ together with the quadrupole components of the deceleration parameter $\hat{\mathfrak{Q}}_o$. The panels showing the shear, in the lower-right part of the plot, are zoomed-in versions of Figure~\ref{fig:shear} to make the contours for the `smooth' simulation more visible. The upper left part of the plot shows the five independent components of the quadrupole of $\hat{\mathfrak{Q}}_o$. As noted earlier, the posterior for the individual quadrupole components in the `smooth' simulation, when measured at low redshift (orange contours), is compatible with zero at $1\,\sigma$ in all cases. The large value of $C_\ell$ seen in Figure~\ref{fig:Cls} therefore needs to be interpreted as a projection effect from five-dimensional parameter space. The five panels highlighted in light green in the lower left part of Figure~\ref{fig:quadrupoles} show the degeneracy between $\mathfrak{H}_o$ and $\hat{\mathfrak{Q}}_o$ in the anisotropic distance--redshift relation. The degeneracy is considerably weaker in the `smooth' simulation where the quadrupole of $\mathfrak{H}_o$ can be measured very robustly. In the \textit{UNITY} simulation the degeneracy remains significant even at high redshift. Using prior knowledge about the expected smallness of the deceleration quadrupole might therefore yield a more robust constraint.

\begin{figure*}
    \centering
    \includegraphics[width=\textwidth]{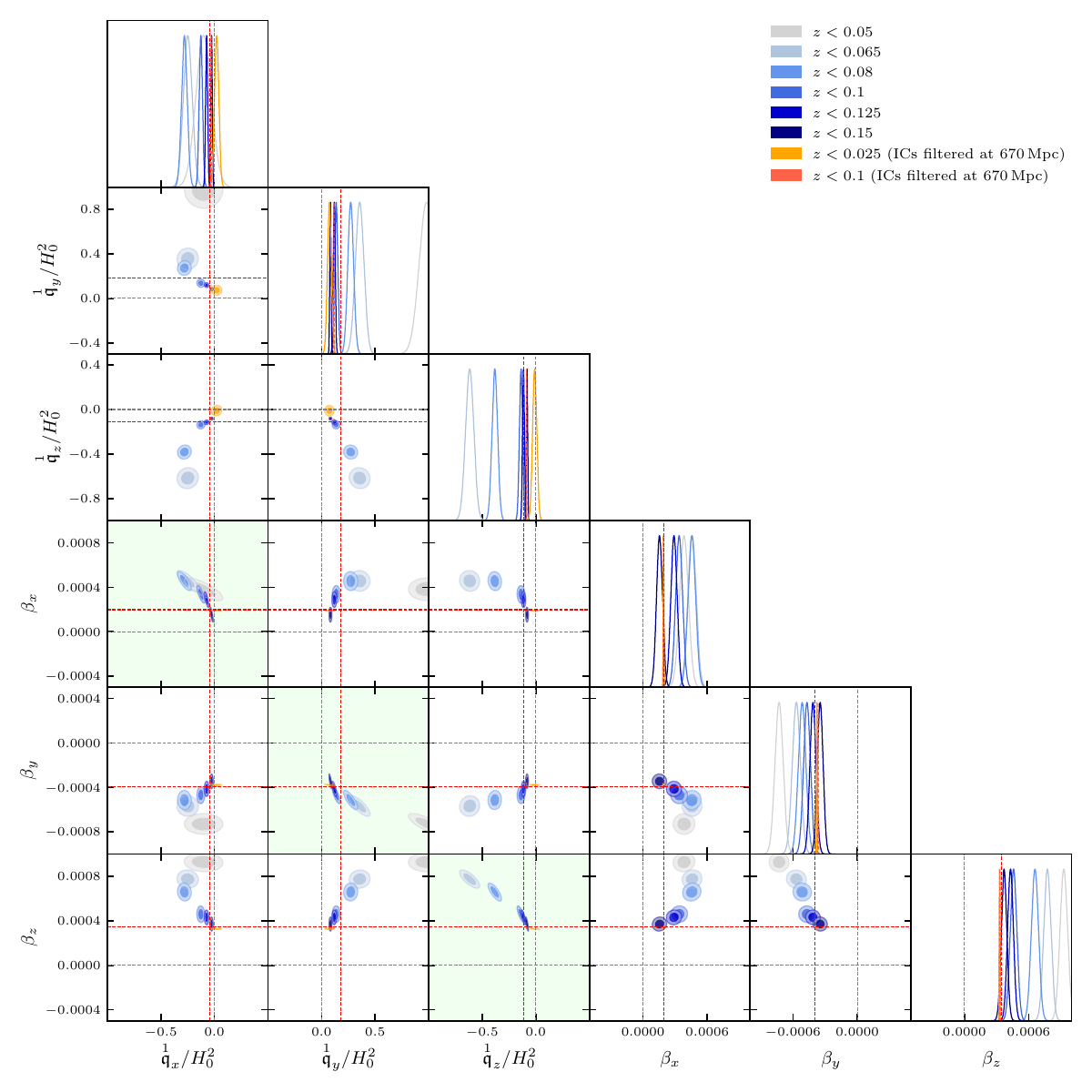}
    \caption{2D marginalised posteriors of dipolar contributions to the distance--redshift relation for different fitting ranges in redshift $z$ (different colours). The sampling is roughly uniform in observed distance $d_A$ to the sources. The three components of $\overset{1}{\mathfrak{q}}_\mu$ characterise the dipole anisotropy of the deceleration parameter whereas the three components of the velocity vector $\boldsymbol{\beta}$ parameterise the peculiar velocity of the observer with respect to the fitted coarse-grained matter congruence. Results from a simulation where perturbations below the scale of 670\,Mpc were filtered out in the initial data are shown in orange and red. For this case, the red dashed lines show an alternative measurement using the properties of null geodesics at the observer location ($z \rightarrow 0$). All shown quantities would be zero in an exact FLRW model (grey dashed lines). The panels highlighted in light green show the degeneracies between pairs of parameters that affect the same dipole axis.}
    \label{fig:dipole}
\end{figure*}
In Figure~\ref{fig:dipole} we show two-dimensional constraints on the components of the dipolar contributions to the distance--redshift relation. We assume from the outset that the expansion rate has vanishing dipole as it would be proportional to the acceleration at the observer, see Eq.~\eqref{eq:Eevolution}. Therefore, the possible sources of a dipole anisotropy are the dipole in the deceleration parameter, $(\overset{1}{\mathfrak{q}}_x,\overset{1}{\mathfrak{q}}_y,\overset{1}{\mathfrak{q}}_z)$, and the observer velocity with respect to the best-fit congruence, $(\beta_x,\beta_y,\beta_z)$. As previously, with increasing maximum redshift we generally see 
the contours for the deceleration anisotropy progress towards the isotropic limit. This progression is much slower for the observer boost, which is still very far away from zero even for the largest redshift interval that goes to $z = 0.15$. For that case, the posteriors of the components of $\boldsymbol{\beta}$ overlap with the ones measured in the `smooth' simulation. For each of the three dipole axes, there is a noticeable degeneracy between the amplitude of the deceleration dipole and the observer velocity, as can be seen by the fact that the corresponding contours are tilted. The degeneracy is largest at low redshift and gradually disappears as the redshift interval grows --- this is expected since the two effects act differently across redshift. As a guide to the eye, we have highlighted the three relevant panels in light green.

The dashed red lines in Figure~\ref{fig:dipole} indicate (in the relevant panels) the deceleration dipole measured in the `smooth' simulation using numerical derivatives at the observer location or the relative velocity of the observer frame with respect to the matter at that same location. The MCMC analysis recovers the observer boost very accurately from both redshift windows (extremely tight orange and red contours), whereas the dipole of the deceleration is not measured as well, especially when using only data at low redshifts $z < 0.025$.

\section{Conclusion}
\label{sec:conclusion}

Based on the cosmographic framework by \citet{Heinesen:2020bej} for analysing the distance--redshift relation in  a model-independent way, we have developed a new approach to constrain the anisotropy in the expansion rate, deceleration, and higher-order differentials of the distance--redshift relation in the real Universe that is inhomogeneous on small scales. These new observables encapsulate information about the inhomogeneities within the local volume over which the data are taken. 
The observational selection function introduces an effective averaging procedure, and as the probed volume increases we see a trend towards the FLRW limit, as expected in the simulated $\Lambda$CDM model of our analysis. 
The departure from FLRW as a function of selection window can be seen as a new observable that can be predicted for any cosmological model, opening up a new way to test and constrain different possible scenarios. Our analysis can be applied directly to real observations.

Here we used synthetic data, generated with $\Lambda$CDM $N$-body simulations with 
relativistic ray tracing, to develop our analysis. Since a third-order Taylor series expansion of the distance--redshift relation (including expansion rate, deceleration, and jerk) has an error larger than $1\%$ already at redshift $z = 0.2$, we employ a Pad\'e approximant that is better behaved over an extended redshift range and limits our truncation error below $0.2\%$. We find that the lowest multipoles in the expansion rate and deceleration can be detected at high significance with a few thousand sources that sample the full sky and have a scatter that is dominated by peculiar motion. Under such optimal measurement conditions, the observed deviation from FLRW is still significant even at redshift $z = 0.15$, i.e.\ in a volume measuring more than 1\,Gpc in diameter. In particular, the variation of the expansion rate can be as large as $1\%$ across the sky, while the deceleration parameter can have variations of order $10\%$. Results for different redshift ranges are summarised in Figure~\ref{fig:Cls}; $C_\ell$ effectively measures the variance of the expansion rate or deceleration across the sky, hence $C_\ell^{1/2}$ indicates the relative fluctuation amplitude. 

In relation to this, it is worth noting that deviations from the $\Lambda$CDM 
initial conditions could change the quoted amplitudes of the anisotropies at various scales. If such deviations happen in our
Universe, as some empirical observations of large-scale configurations and flows of matter may indicate \citep[see, e.g.,][]{Migkas:2021zdo,Peebles:2023jzn,Watkins:2023rll,vonHausegger:2024jan}, this could lead to measurements of multipoles that differ quantitatively in amplitude from those of the fiducial observer of this analysis. If we are atypical observers in one way or another, this can also cause differences in the measured data relative to the fairly typical $\Lambda$CDM observer of our analysis.  
The methods that we have presented in this work can deal with both of these scenarios: it is precisely the strength of the cosmography framework presented here that it allows to be agnostic about the particularities of the ambient geometry and perturbation statistics.

We compare our new analysis to previous work where the multipoles of the distance--redshift relation are computed directly from the kinematical properties of the matter congruence at the observer location. This requires a unique notion of matter congruence that only exists in a Universe where the overlap of multiple streams of matter (often called `shell-crossing') is strictly avoided. We therefore run a simulation using initial conditions where small-scale perturbations were artificially removed by applying a low-pass filter. We find that the local definition of the expansion rate agrees well with our new analysis when applied to the same simulation. The agreement is indeed extremely precise for the monopole and quadrupole of the expansion rate, where both methods give the same departure from the FLRW limit. The inferred deceleration parameter and its multipoles, when measured using our new analysis, do not resemble the local measurement that well. This may be related to the fact that 
the latter measurement is based on numerical derivatives of data, which are prone to numerical issues --- such as particle aliasing when performing a particle-to-mesh projection --- especially for higher derivative orders.
We also note, however, that our MCMC analysis is based on a Gaussian likelihood model, which does not well describe the true scatter of the data when small-scale structure (otherwise the primary cause of scatter) has been artificially suppressed. 

While finalising this article, an interesting study by \citet{Kalbouneh:2024szq} appeared that is in many ways complementary to our work. Using a toy model for a large inhomogeneity in the local Universe that still admits a unique matter congruence, the anisotropies in the distance--redshift relation are discussed in the cosmographic framework. Their discussion of the role of the observer frame is consistent with our approach, showing that it can be included in the analysis as we both suggest. While they advocate using a Taylor series expansion in redshift for the distance--redshift relation \citep[similar to earlier work, including some of our own, e.g.][]{Heinesen:2020bej,Heinesen:2021azp}, we believe that using the inverse series, i.e.\ an expansion in angular diameter distance, has advantages when dealing with real data.\footnote{In practice, there may be subtleties in using the distance as a parameter, since the distances to astrophysical sources are often not \textit{a priori} known and must be fitted together with the cosmology, as is for instance the case in the  standardisation of supernovae of type 1a by \citet{Tripp:1997wt}.} 
In particular, angular diameter distance is a smooth parameter along any null ray and is in practice strictly monotonic over a significant portion of the local light cone. The irreducible, intrinsic scatter of the data is instead linked mainly to Doppler contributions that affect the observed redshift, rendering this quantity discontinuous and badly non-monotonic. Our approach does not require a binning in redshift and can effortlessly incorporate negative redshifts which could be relevant for accurately measuring large observer boosts (it is of course physically impossible to find negative values of the angular diameter distance of a source).

In the realistic case where highly nonlinear structures form in the Universe, the notion of a matter congruence becomes ambiguous. Our approach provides an observational procedure to construct an effective congruence for a given survey footprint, allowing us to quantify the expected deviations from the FLRW limit. Since these deviations could be measured out to very large scales, they could be used as new probes in the analysis of cosmic large-scale structure, shedding a new light on the questions of how data should be interpreted in an inhomogeneous Universe and how exactly the FLRW limit is approached when taking summary statistics.

\begin{acknowledgments}
The simulations used in this work were originally run on the DiRAC DIaL system, operated by the University of Leicester IT Services, which forms part of the STFC DiRAC HPC Facility (\url{www.dirac.ac.uk}). Reusing existing simulation data greatly reduces the carbon footprint of our research activity. JA, RD and MK acknowledge financial support from the Swiss National Science Foundation. Support for HJM was provided by NASA through the NASA Hubble Fellowship grant HST-HF2-51514.001-A awarded by the Space Telescope Science Institute, which is operated by the Association of Universities for Research in Astronomy, Inc., for NASA, under contract NAS5-26555. CC is supported by the UK Science \& Technology Facilities Council Consolidated Grant ST/T000341/1.
\end{acknowledgments}

\appendix

\section{Comparison with the \texttt{Einstein Toolkit}}\label{appx:ETcomp}

In this Appendix, we consider two simulations with identical initial data performed with different codes: \texttt{gevolution} and the \texttt{Einstein Toolkit}\footnote{\url{einsteintoolkit.org}} \citep[ET;][]{Loffler:2011ay,Zilhao:2013hia}. 
The purpose of this test is to ensure that our calculations of the cosmographic parameters using $N$-body simulation data are consistent with a simulation of a strict fluid congruence in a carefully controlled setup, which we describe below.

\begin{figure*}[t]
     \centering
     \begin{minipage}[b]{0.49\textwidth}
         \centering
         (a)
         \includegraphics[width=\textwidth, trim= 5cm 0cm 5cm 0cm, clip]{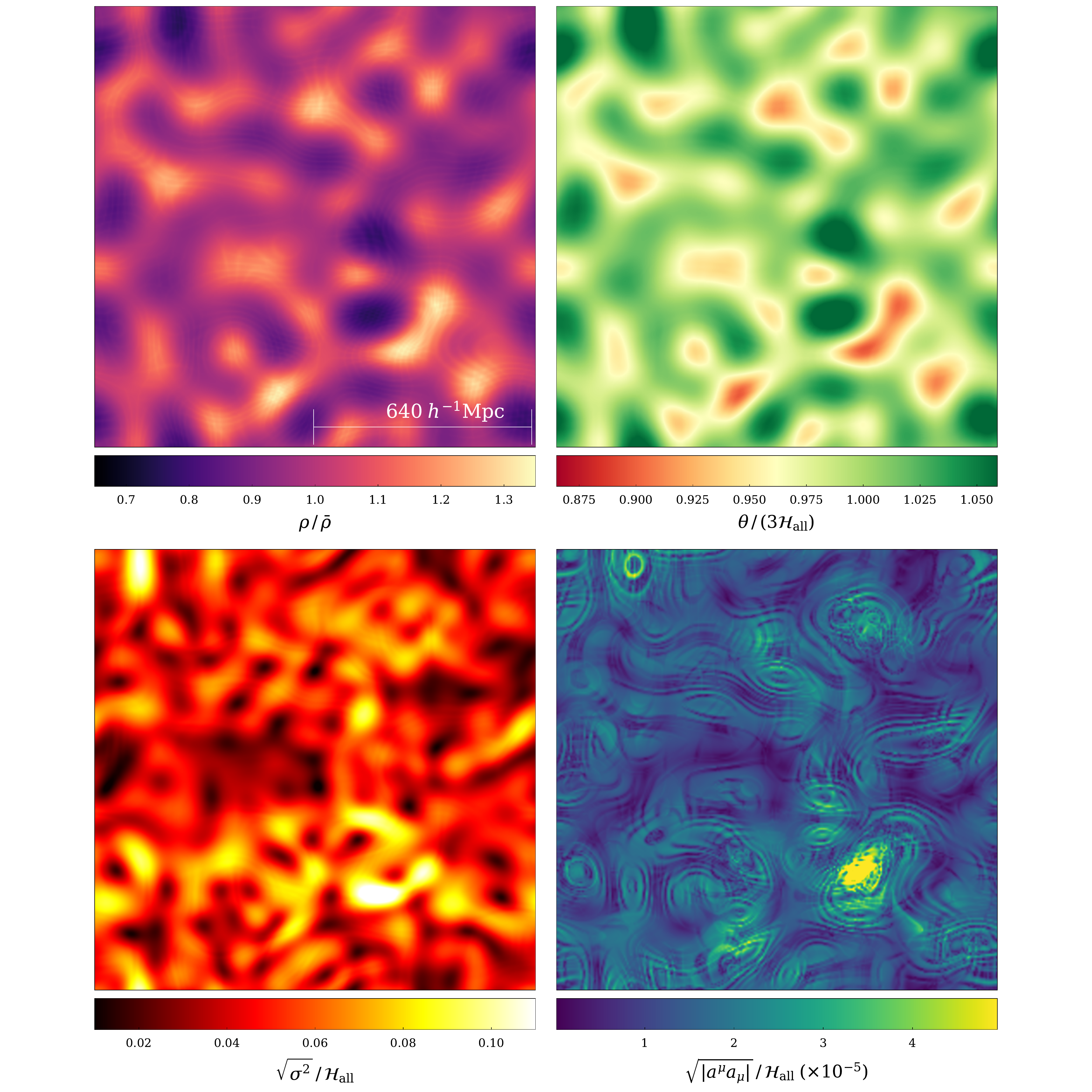}
     \end{minipage}
     \hfill
     \begin{minipage}[b]{0.49\textwidth}
         \centering
         (b)
         \includegraphics[width=\textwidth, trim= 5cm 0cm 5cm 0cm, clip]{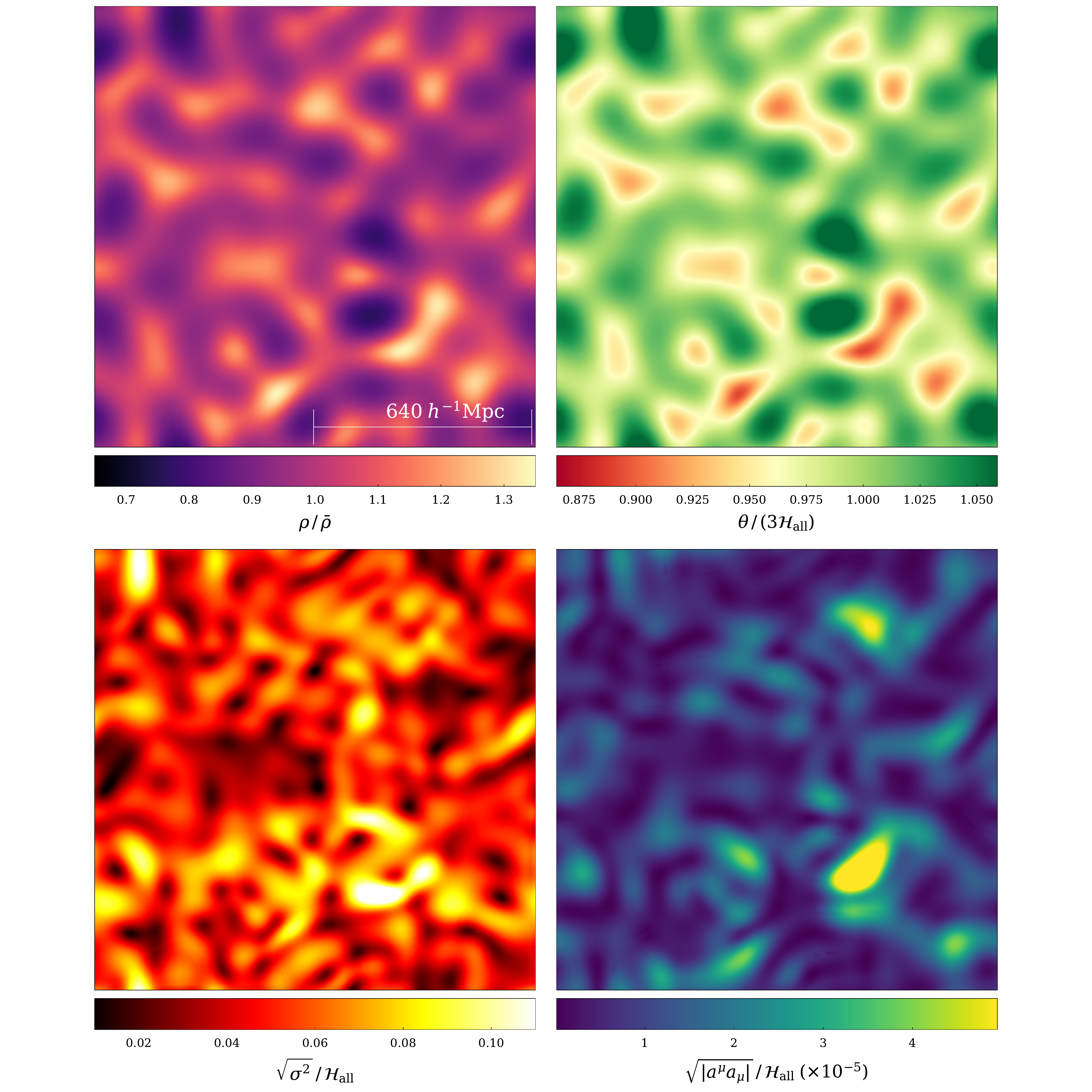}
     \end{minipage}
        \caption{2D slices at redshift $z=0$ of the density, scalar expansion, shear scalar, and acceleration scalar (top-left to bottom-right, respectively) for \texttt{gevolution} data (a) and ET data (b) for runs with identical initial conditions. The physical box length is $1.28\,h^{-1}\,\mathrm{Gpc}$ and all modes with wavelength below $200\,h^{-1}\,\mathrm{Mpc}$ were removed from the initial conditions.}
        \label{fig:2Dplots}
\end{figure*}

The simulation performed with the ET mimics the setup used in \citet{Macpherson:2018btl,Macpherson:2021gbh}, namely that of a matter-dominated universe initialised as a linearly-perturbed Einstein-de Sitter (EdS) model. The initial conditions are set up with \texttt{FLRWSolver}\footnote{\url{https://github.com/hayleyjm/FLRWSolver_public}} \citep{Macpherson:2016ict} as Gaussian random fluctuations following the matter power spectrum at $z_\mathrm{ini}=1000$ from \texttt{CLASS}\footnote{\url{http://class-code.net}} \citep{Blas:2011rf}.
To replicate the same spacetime solution with \texttt{gevolution} we
generate the initial data for the ET simulation --- which adopts a fluid approximation for the matter content --- and then translate these to particle displacement fields and velocities for the $N$-body ensemble of \texttt{gevolution} using a method described in \citet{Adamek:2021rot}.

The matter power spectrum we use to generate the initial data has initial perturbations smaller than 
$\sim 200\, {\rm Mpc}/h$ removed, namely, we set $P(k>k_{\rm max})=0$ for $k_{\rm max}=2\pi / \lambda_{\rm min}$ and $\lambda_{\rm min}=200\, {\rm Mpc}/h$. Then the power at $k<k_{\rm max}$ is set according to the output power spectrum from \texttt{CLASS}. 
We choose this particular scale cut to ensure that the structures appear at large scales
only, and thus should remain well within the perturbative regime where the fluid description is valid. For this test, we choose an initial smoothing scale which is distinct from the smoothed simulations used in the main text in order to replicate the initial data of an existing ET simulation.
In the presence of nonlinear structures, we might expect different results when comparing simulations performed under a fluid approximation and with an $N$-body description, especially when sampling small enough scales such that we resolve shell-crossing, in which case the fluid approximation breaks down completely. 

Both simulations use a mesh of $256^3$ points, and the $N$-body
simulation is run with $1024^3$ particles to reduce noise from the particle discretisation. We choose a physical box length of $L=1.28\,\mathrm{Gpc}/h$ for both simulations. 
The simulations are both initialised in the Newtonian gauge, and the lapse and shift chosen in the ET simulation will maintain this gauge so long as the perturbations remain in the linear regime \citep[see][for more details on the gauge used]{Macpherson:2018btl}. The relativistic $N$-body simulation with \texttt{gevolution} uses the Newtonian gauge throughout.

Figure~\ref{fig:2Dplots} shows two-dimensional slices of the fluid scalars in the final snapshots of the two simulations run with \texttt{gevolution} (left panels) and the ET (right panels). For each simulation, we show the density field, expansion scalar, shear scalar, and acceleration magnitude (top-left to bottom-right on each side, respectively). 
While the outputs of the two simulations show a good overall agreement on large scales, the particle-to-mesh projection used to construct the fields from the $N$-body ensemble leads to small-scale numerical artefacts that are particularly noticeable in the bottom-right subplot showing the acceleration.

The smooth density and velocity fields constructed from \texttt{gevolution} were then passed through the same
post-processing code as for the ET data \citep[\texttt{mescaline}; see][]{Macpherson:2018btl}. 
For both simulations, we calculate the effective cosmological parameters on the slice at redshift $z=0$ 
\citep[see][for further details on the calculation itself]{Macpherson:2021gbh}.
We calculate the effective Hubble, deceleration, jerk, and curvature parameters first for a set of 100 observers randomly placed in each simulation domain (with equivalent positions between simulations), with lines of sight drawn in the direction of $12\times N_{\rm side}^2$ \texttt{HEALPix} pixels with $N_{\rm side}=32$.

\begin{figure*}[t]
     \centering
     \begin{minipage}[b]{0.49\textwidth}
         \centering
         (a)
         \includegraphics[width=\textwidth, trim= 0.5cm 0cm 0.5cm 0cm, clip]{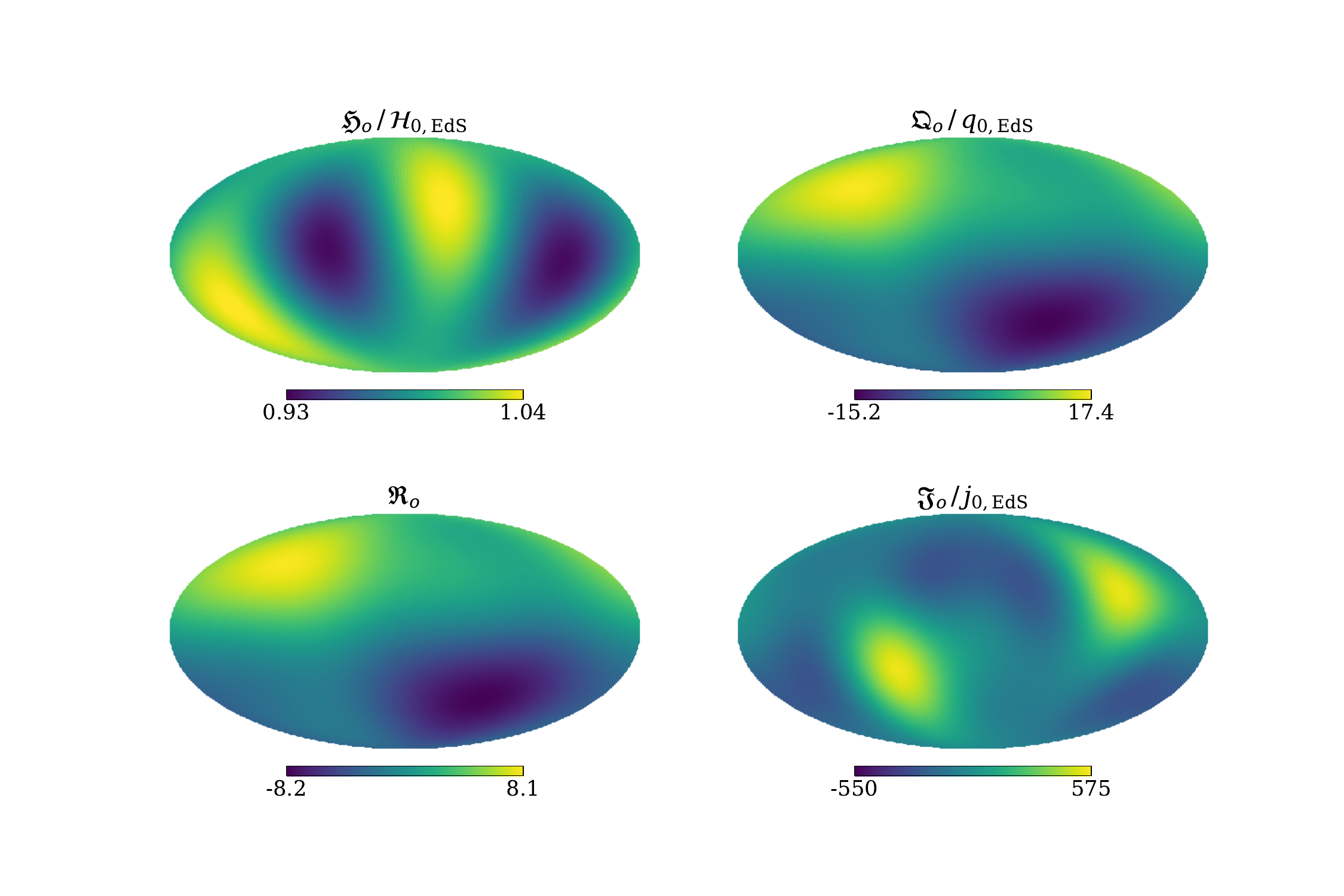}
     \end{minipage}
     \hfill
     \begin{minipage}[b]{0.49\textwidth}
         \centering
         (b)
         \includegraphics[width=\textwidth, trim= 0.5cm 0cm 0.5cm 0cm, clip]{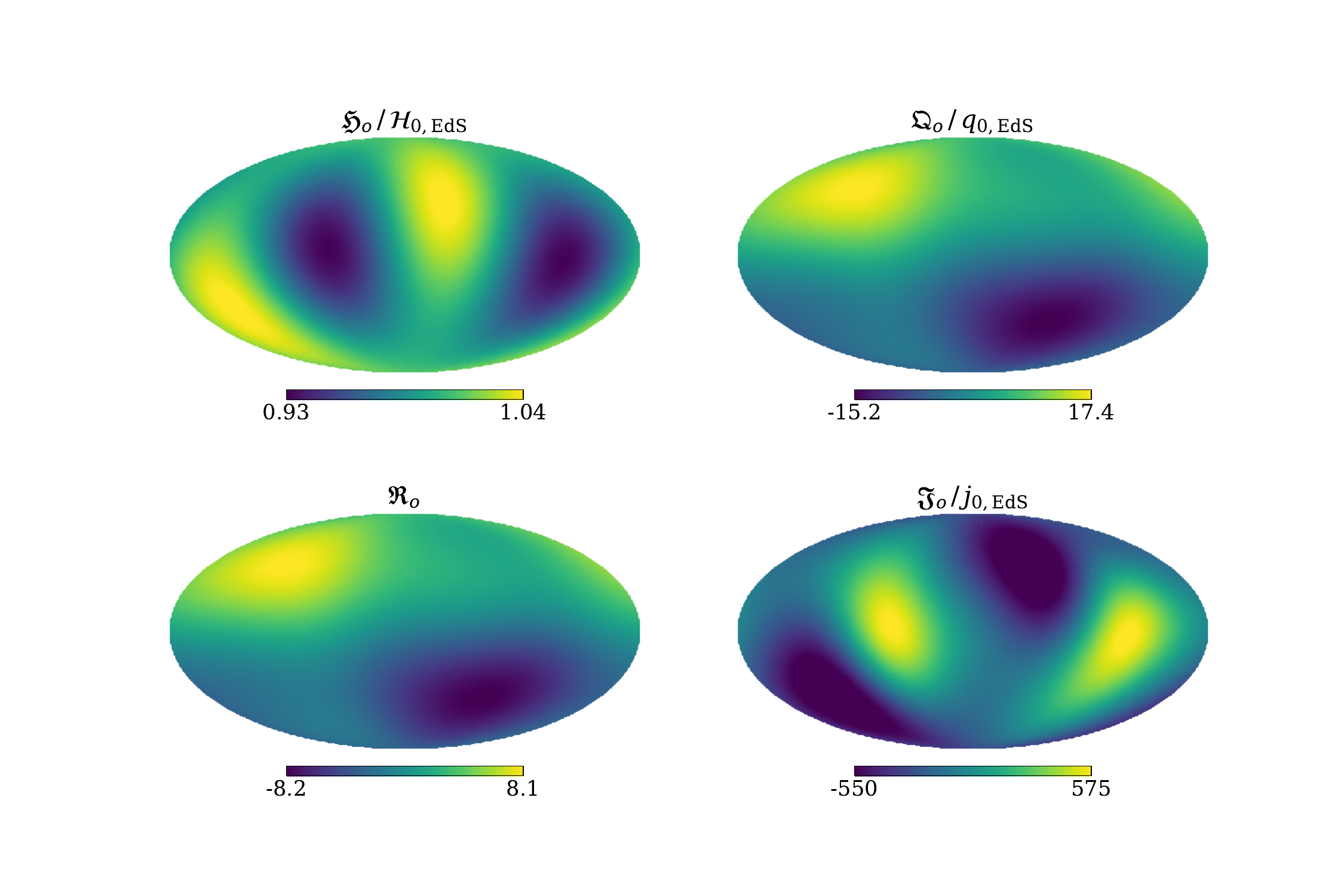}
     \end{minipage}
        \caption{Sky maps of the effective parameters for the same observer in the simulation from \texttt{gevolution} (a) and ET (b) with domain size $1.28\,h^{-1}\,\mathrm{Gpc}$ and all power beneath $200\,h^{-1}\,\mathrm{Mpc}$ removed from the initial conditions. Each parameter is shown normalised by its EdS counterpart, with the exception of curvature.}
        \label{fig:skymap_200Mpc}
\end{figure*}

Figure~\ref{fig:skymap_200Mpc} shows a set of Mollweide projections of the effective Hubble, deceleration, curvature, and jerk (top-left to bottom-right, respectively, in each set of panels) for one observer placed at the same location in the simulations from \texttt{gevolution} (left panels) and ET (right panels). With the exception of curvature, all parameters are normalised by their EdS counterpart. We notice both the anisotropic signatures and amplitudes of the parameters are very close between the simulations, with the exception of the jerk parameter (bottom right in both sets of panels), which shows visible differences in anisotropic signature and a larger difference in amplitude.

Since this analysis is based on numerical derivatives of the fields at the observer location, we expect that high-order derivatives --- like the ones that enter the calculation of the jerk --- may become contaminated by the particle discretisation, possibly requiring a higher-order particle-to-mesh projection to mitigate (we used `cloud-in-cell' projection). However, we do not expect that our MCMC analysis in the main text suffers from similar shortfalls, as no derivatives of data are taken at all in this case.

Next, we consider a set of 1000 observers (again placed in identical positions between the two simulations) each with 300 randomly-drawn lines of sight (lines of sight are unique for each observer but identical between simulations). For each line of sight we calculate the four effective cosmological parameters, and average the result over the sky for each observer. This procedure was done in \citet{Macpherson:2021gbh} to provide an estimate on the observers ``measurement'' of these parameters, and to assess how they might vary with observer position. 

\begin{figure*}[t]
    \centering
   \includegraphics[width=0.9\textwidth]{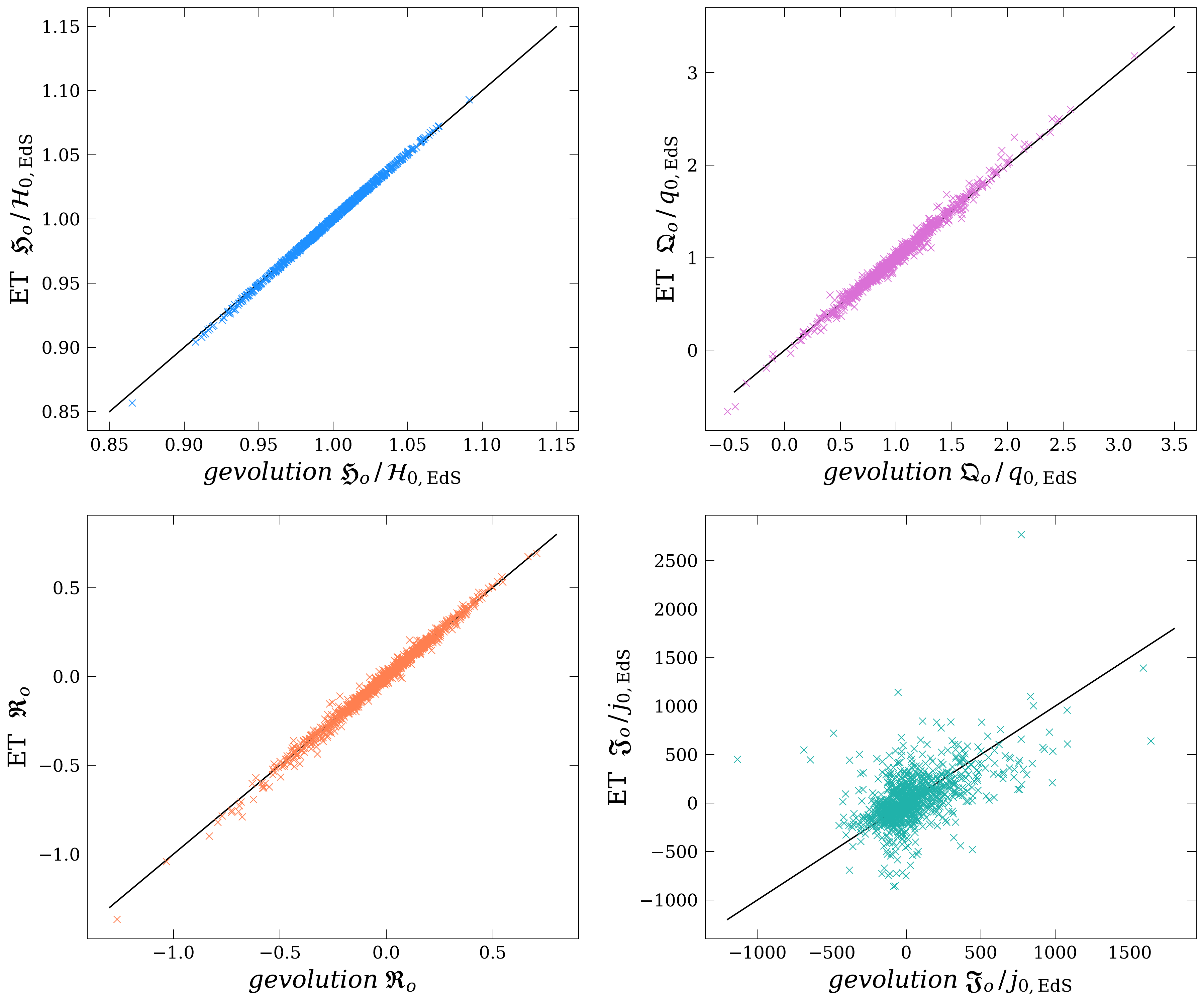}
    \caption{Sky averages for effective cosmological parameters for 1000 observers in  \texttt{gevolution} ($x$-axes) and ET ($y$-axes). Each point shows the effective parameters relative to the EdS value for one observer as averaged over 300 randomly drawn lines of sight. Solid black lines in each panel indicate a 1:1 match, i.e.\ a perfect agreement between ET and \texttt{gevolution}.}
    \label{fig:randsky_200Mpc}
\end{figure*}

Figure~\ref{fig:randsky_200Mpc} shows these sky averages of the four effective cosmological parameters (panels; normalised by each EdS counterpart) for all observers in the simulations from \texttt{gevolution} ($x$-axes) and ET ($y$-axes). Solid black lines in each panel shows a 1:1 correspondence, i.e.\ what we would expect if the ET and \texttt{gevolution} results were identical. The effective Hubble parameter, deceleration parameter, and curvature parameter show a good agreement between the codes.
We note there are some small differences, however, we emphasise that we are comparing two independent codes which have different numerical errors as well as distinct descriptions of the matter field itself. As in Figure~\ref{fig:skymap_200Mpc} and for reasons already discussed, we do not find a good agreement for the effective jerk parameter.

We conclude that the simulation run with \texttt{gevolution} provides a very good description of the large-scale congruence in this controlled situation in which we expect it to exist.

\bibliography{references}{}
\bibliographystyle{aasjournal}

\widetext

\end{document}